\documentclass[aps,pre,showpacs,amsmath,amsfonts,amssymb,superscriptaddress,twocolumn,nofootinbib]
{revtex4-1}

\usepackage[english]{babel}
\usepackage[pdftex]{graphicx}
\usepackage{tikz}
\usepackage{rotating}
\usepackage[pdftex,hypertexnames=true,linktocpage=true]{hyperref} 
\hypersetup{colorlinks=true,linkcolor=blue,anchorcolor=blue,citecolor=magenta,filecolor=blue,urlcolor=blue,bookmarksnumbered=false,pdfview=FitB}

\usepackage{amsthm}
\usepackage{framed}
\usepackage{amssymb}
\usepackage{amsmath,mathrsfs}
\usepackage{hhline}
\usepackage{ulem}
\usepackage{fancyhdr}
\usepackage{simplewick}

\everymath{\displaystyle}
\everymath{\displaystyle}

\everymath{\displaystyle}
\newcommand{\beq}{\begin{equation}}
\newcommand{\eeq}{\end{equation}}

\newcommand{\bi}{\begin{itemize}}
\newcommand{\ei}{\end{itemize}}
\newcommand{\D}{\mathrm{d}}

\newcommand{\affA}{Aix-Marseille University, Marseille, France}
\newcommand{\affB}{CNRS Centre de Physique Th\'eorique UMR7332,
13288 Marseille, France}
\newcommand{\affC}{Centre d'Immunologie de Marseille-Luminy,
13288 Marseille, France}
\newcommand{\affD}{CNRS, UMR7280, Marseille, France}
\newcommand{\affE}{INSERM, U1104, Marseille, France}
\newcommand{\affG}{ Department of Chemistry, Rice University, Houston, USA}

\begin{document}


\title{Experimental detection of long-distance interactions between biomolecules through their diffusion behavior: Numerical study}

\author{Ilaria Nardecchia}
\email{i.nardecchia@gmail.com}
\affiliation{\affB}\affiliation{\affC}\affiliation{\affD}\affiliation{\affE}
\author{Lionel Spinelli}
\email{spinelli@ciml.univ-mrs.fr}
\affiliation{\affC}\affiliation{\affD}\affiliation{\affE}
\author{Jordane Preto}
\email{jordane.preto@gmail.com}
\affiliation{\affG}
\author{Matteo Gori}
\email{gori6matteo@gmail.com}
\affiliation{\affB}\affiliation{\affA}
\author{Elena Floriani}
\email{floriani@cpt.univ-mrs.fr}
\affiliation{\affB}\affiliation{\affA}
\author{Sebastien Jaeger}
\email{sebjaeger@gmail.com}
\affiliation{\affC}\affiliation{\affD}\affiliation{\affE}
\author{Pierre Ferrier}
\email{ferrier@ciml.univ-mrs.fr}
\affiliation{\affC}\affiliation{\affD}\affiliation{\affE}\affiliation{\affA}
\author{Marco Pettini}
\email{pettini@cpt.univ-mrs.fr}
\affiliation{\affB}\affiliation{\affA}

\begin{abstract}
The dynamical properties and diffusive behavior of a collection of mutually 
interacting particles are numerically investigated for two types of long-range 
interparticle interactions: Coulomb-electrostatic and dipole-electrodynamic.
It is shown that when the particles are uniformly 
distributed throughout the accessible space, the self-diffusion coefficient is 
always lowered by the considered interparticle interactions, irrespective of
their attractive  or repulsive character. This fact is also confirmed by a simple
model to compute the correction to the Brownian diffusion coefficient due to 
the interactions among the particles. These interactions are also responsible 
for the onset of dynamical chaos and an associated chaotic diffusion which still
follows an Einstein-Fick like law for the mean square displacement as a function of time.
Transitional phenomena are observed for Coulomb-electrostatic (repulsive) and 
dipole-electrodynamic (attractive) interactions considered both separately and
in competition. The outcomes reported in this paper clearly 
indicate a feasible experimental method to probe the activation of resonant 
electrodynamic interactions among biomolecules.
\end{abstract}

\date{\today}
\maketitle
\section{Introduction}\label{Intro}
The present work is the follow-up of a recent paper of ours \cite{Preto2012},
where a first step was made to investigate why and how long-range intermolecular
interactions of electrodynamic nature might influence the $3$D encounter dynamics of biological partners.
Based on a simple analytical study in one spatial dimension, we have reported
quantitative and qualitative dynamical properties that will stand out in case such interactions play an active role at the biomolecular level. Moreover, non-negligible effects were reported in a parameter domain 
 accessible to standard laboratory techniques suggesting that the contribution of long-range electrodynamic interactions in biological 
processes might be well estimated from experimental measurements.
 The physical observable chosen 
 (the first encounter time between two interacting biomolecules) turns out hardly measurable in practice because it requires to follow the 
dynamics of single molecules. 
Thus the present work aims at filling this gap between theory and experimental
feasibility. This is achieved by investigating some transport properties of long
range interactions acting among a set of particles freely moving in a fluid
environment. 

The novelty of the present work is that one dimensional analytic results in
\cite{Preto2012} are here replaced by $3$D numerical results in a more realistic
context. In fact, biomolecules, which are typically charged, move in three
dimensional space where they are subjected to  several interactions out of which 
there is at least one kind of long-range ones: 
electrostatic interactions. Thus we begin by considering Coulomb interactions,
both screened and unscreened, for which all the
parameters can be precisely assigned. On this basis
we get a reference scenario  allowing an assessment of the sensitivity of
diffusion to forces which are undoubtedly active among charged biomolecules.
Then  we make electrodynamic forces enter the game:
 by studying their possible
competition with Coulomb forces we can find out how new characteristic
features  of the concentration dependence of diffusion can 
emerge making the difference with the previous case. Whence feasible experiments
can be identified.

Now, let us quickly outline the framework of the problem of detecting long
range electrodynamic intermolecular interactions.
The starting point is the observation of the fact that the high efficiency, rapidity and robustness of the complex network of biochemical reactions in living cells must involve directed interactions between cognate partners. 
 This should be especially true for the recruitment of biomolecules at a long distance in order to make them available at the right time and at the 
 right place. A long-standing proposal  \cite{Frohlich1968,Frohlich1977,Frohlich1980} surmises that beyond all the
 well-known short-range forces (chemical, covalent bonding, H-bonding, Van der Waals) biomolecules could interact also at a long distance by means of 
 electrodynamic forces, generated by collective vibrations bringing about large dipole moment oscillations. The existence of collective excitations within 
 macromolecules of biological relevance (proteins and polynucleotides) 
  is well documented experimentally, e.g. through the observation of
  low-frequency vibrational modes in the Raman and far-infrared (THz) spectra
  \cite{Chou1988,Fischer2002,Xie2001,Genzel1976,Urabe1998}.
  These spectral features are commonly attributed to coherent collective oscillation modes of the whole molecule (protein or DNA) or of a substantial fraction of its atoms. These collective conformational
   vibrations are observed in the frequency range of $0.1-10$ THz
   \cite{Markelz2002,Markelz2000,Acbas2014}. A-priori collective excitations can
   be switched on and off by suitable environmental conditions (mainly energy
   supply \cite{Frohlich1968}), a property which is a-priori necessary in a
   biological context.
   Also, they can entail strong resonant dipole interactions between biomolecules when they oscillate with the same pattern of frequencies. 
   Resonance would thus result in selectivity of the interaction. Then the fundamental question is: does Nature exploit these long-distance 
   electrodynamic intermolecular forces in living matter? In other words, are these forces sufficiently strong to play the above surmised role? 
   Note that while electrostatic interactions between charges/dipoles in the cytoplasm are exponentially damped with distance, Debye screening 
   proves generally inefficient for interactions involving oscillating electric fields. The electromagnetic field radiated by charges/dipoles in 
   the cytoplasm oscillating faster than hundreds of MHz is not affected by
   Debye screening \cite{deXammarOro1992,deXammarOro2008} and is able to produce
   long distance interactions.
To answer the questions raised above one has to devise a technologically possible experimental setup {\it in vitro} to begin with - to detect 
some direct physical consequence of the action of long-range interparticle
interactions. As we shall see throughout this paper, long-range interactions
markedly affect the self-diffusion properties of particles. And this is true for electrostatic as well as for electrodynamic interactions, though they entail different phenomenologies with some common features.

By long-range interactions we mean an interaction potential falling off with
the interparticle distance $r$ as $1/r^\nu $ with $\nu \leq d$, $d$
being the spatial dimension of the system. As well, in a looser sense, we also
mean that the interparticle interactions act at a long distance, ``long''
meaning much larger than usual distance for which chemical and Van der Waals
forces act. Hence, in what follows, by ``long distance'' we
mean distances varying from several hundreds to several thousands of 
Angstr\"om.
As we shall see, for collections of solute particles homogeneously distributed
in a given volume, the presence of deterministic forces beside the stochastic
ones (mimicking the collisions of water molecules against a solute
macromolecule) entails a slowing down of diffusion, thus a decrease of the diffusion
coefficient. And this occurs independently of the attractive or repulsive nature
of the interparticle forces.
An independent signature of an increasing strength of the average interparticle
interactions is provided by an increase of the degree of chaoticity of the
dynamics, as measured by the largest Lyapunov exponent.

In Section \ref{Sec-Models} we give the equations of motion of an ensemble of
solute molecules subjected to a random force plus the sum of all the deterministic forces due to
mutual interactions and we  define the three different
intermolecular interactions potentials that we used: Coulomb screened
(short-range repulsive); pure Coulomb (long-range repulsive); dipole-dipole
(long-range attractive) interactions of electrodynamic origin. In the same
Section, we also propose a simple theoretical derivation of a formula that
accounts for a correction to the Brownian diffusion coefficient in presence of interactions among the solute molecules.

In Section \ref{NumRes} we report the outcomes of the numerical study of the
previously mentioned models and we comment on the observed phenomenology.

The Section \ref{Sec-concluding} is devoted to some concluding remarks about 
the results presented throughout the present work. Moreover, 
for what concerns the 
feasibility of laboratory experiments aimed at detecting long-range interactions
among biomolecules, we have identified an observable - the self-diffusion
coefficient- which can be easily accessed with available experimental techniques
and which is very sensitive to intermolecular deterministic interactions.

\section{Models}\label{Sec-Models}
 
In the present Section we define the model equations, the molecular interaction potentials, the numerical 
algorithm and the relevant observables for the numerical study of an ensemble of mutually interacting particles 
in presence of an external random force.
   
\subsection{Basic equations}\label{BasicEq}

We consider a system composed of $N$ identical molecules, modeled as spherical
Brownian particles of radius $R$, mass $M$ and a net number of electric 
charges $Z$, moving in a fluid with viscosity
 $\eta$ at a fixed temperature $T$, interacting through a pairwise potential
 $U(r)$ which depends only on the distance $r$ between their centers.\\
Under the assumption that the friction exerted by the fluid environment on the
particles is described by Stokes' law,  the dynamics of the system is given by
$N$ coupled Langevin equations \cite{Gardiner2009}:
\begin{equation}
\begin{split}
{M} \frac{d^{2} \boldsymbol{r}_{i}}{dt^{2}}  = & - \gamma \frac{d \boldsymbol{r}_{i}}{dt}  - \sum\limits_{j=1,j\neq i}^{N} \boldsymbol{\nabla}_{\boldsymbol{r}_i} U\left(|\boldsymbol{r}_{i}-\boldsymbol{r}_{j}|\right)
+\\ &+\sqrt{2 \gamma k_{B} T} \boldsymbol{\xi}_i(t) \qquad \text{for } i=1,...,N
\label{lang.}
\end{split}
\end{equation}
where $\boldsymbol{r}_{i}$ is the coordinate of the center of i-th particle, $\gamma=6\pi\eta R$ is the friction coefficient and $k_{B}$ is the Boltzmann constant. 
The stochastic displacements are uncorrelated so that
$\boldsymbol{\xi}(t)=(\boldsymbol\xi_1,...,\boldsymbol\xi_N)$ is a
$3N$-dimensional random process modeling the fluctuating force due to the
collisions with water molecules, usually represented as a Gaussian white noise
process satisfying:
\begin{equation}
\begin{cases}
\langle \xi^{\alpha}(t)\rangle_{\xi}=0\\
\\
\langle \xi^{\alpha}_{i}(t) \xi^{\beta}_{k}(t')\rangle_{\xi} =
\delta^{\alpha\beta} \delta_{ik} \delta(t-t')\\
\end{cases}
\label{whitenoise}
\end{equation}
where $\alpha,\beta=x,y,z$ are the cartesian
components of $\boldsymbol{\xi}_{i}$'s
and $\langle \cdot \rangle_{\xi}$ stands for an average
over many realizations of the noise process. As the
random process is stationary the average over different 
realizations of the noise is equivalent to a time average
\begin{equation}
\langle f(\boldsymbol \xi)\rangle_{\xi}=\lim_{t\rightarrow+\infty}\dfrac{1}{t}\int_{0}^{t}f(\boldsymbol \xi(\tau))\D \tau=\lim_{t\rightarrow +\infty}\langle f(\boldsymbol \xi(t)) \rangle_{t}
\label{statnoise}
\end{equation}
Considering times much larger than the relaxation time $\tau_{r}=M /\gamma$,
we can neglect inertial effects obtaining the overdamped limit for Eqs.
\eqref{lang.}:
\begin{equation}
\begin{split}
&\gamma \frac{d \boldsymbol{r}_{i}}{dt} = - \sum\limits_{j=1, j \neq i}^{N}\boldsymbol{\nabla}_{\boldsymbol{r}_i} U\left(|\boldsymbol{r}_{i}-\boldsymbol{r}_{j}|\right)
+\\
&+\sqrt{2 \gamma k_{B}T} \boldsymbol{\xi}_{i}(t) \qquad i=1,...,N
\end{split}
\label{lang2.}
\end{equation}
In systems like the one we are interested in (involving protein or nucleid acids
in aqueous medium) $\tau_{r}$ is negligible compared with the characteristic
time scales for experimental observations \footnote{I.e. for a biomolecule with a hydrodynamic radius $R=2\times 10^{-3} \mu\mathrm{m}$ and mass $M=15 K\text{Da}$ in pure water at $300\text{K}$,
the relaxation time $\tau_{r}$ is in the order of $10^{-6}\mu \text{s}$.},
so we can assume that the dynamics for such systems is described by Eqs. \eqref{lang2.}.\\
As the deterministic interactions are in general non linear,
we are dealing with a system of first order Stochastic Differential Equations (SDEs) which describes a 
randomly perturbed nonlinear N-body dynamical system  with
an expected complex (chaotic) dynamics since the integrability is exceptional.
For this reason, we undertake the numerical integration
of Eqs.\eqref{lang2.}. 
We remark that Eqs.\eqref{lang2.} can be considered as a Lagrangian description
of a system whose Eulerian description is given by a Fokker-Planck equation for
the N-body probability distribution
$P_N(\boldsymbol{r}_1,...\boldsymbol{r}_N,t)$ \cite{Chavanis2011} of the form:
\begin{equation}
\begin{split}
&\dfrac{\partial P_N}{\partial t}=\gamma\sum_{i=1}^N \boldsymbol \nabla_{\boldsymbol r_i}\cdot\Biggr(
D_B \boldsymbol \nabla_{\boldsymbol r_{i}}P_N+P_N \dfrac{ \boldsymbol \nabla_{\boldsymbol r_i} U(\boldsymbol r_1,...,\boldsymbol r_n)}{\gamma}\Biggr)
\label{FokkerPlanck}
\end{split}
\end{equation}
where $D_B=k_{B}T/\gamma$ is the Brownian diffusion coefficient and
$U(\boldsymbol r_1,...,\boldsymbol r_n)= \sum_{i=1}^N \sum_{j>i}U(|\boldsymbol r_i-\boldsymbol r_j|)$
is the total interaction energy.
It is well known  that Gibbs configurational distribution
$P_N^{eq}=P_N^{eq}(\boldsymbol{r}_1,...,\boldsymbol{r}_N)$ is the stationary solution of Eq.\eqref{FokkerPlanck} which also minimizes free energy \cite{Chavanis2011}: 
\begin{equation}
P_N^{eq}=\dfrac{1}{Z}\exp\left[-\beta U(\boldsymbol r_1,...,\boldsymbol r_n)\right]
\label{gibbsdistr}
\end{equation}
where $\beta=1/k_{B}T$ and
\begin{equation}
Z=\int \exp\left [-\beta U(\boldsymbol r_1,...,\boldsymbol r_n)\right]\prod_{i=1}^{N}\D\boldsymbol r_{i}
\end{equation}
The distribution of Eq.\eqref{gibbsdistr} defines an equilibrium measure
$\mu^{eq}$
\begin{equation}
\mu^{eq}(f(\boldsymbol r_i))=\int f(\boldsymbol r_i)
P^{eq}_N(\boldsymbol r_i)\prod_{i=1}^{N}\D\boldsymbol r_{i}
\label{gibbs_measure}
\end{equation}
which is invariant respect to the flow defined by Eqs. \eqref{lang2.}.
As we are interested especially in the behavior of systems described by
Eqs. \eqref{lang2.} in the limit $t\rightarrow +\infty$, we assume that
the system thermalizes without any dependence on initial conditions,
\textit{i.e.} for every initial configuration
$\{\boldsymbol{r}_i(0)\}_{i=1,..N}$ it exists a time $\tilde{t}$ such as
$P_N(t)\simeq P^{eq}_N$ for $t>\tilde{t}$.

\subsection{Model potentials}\label{AnalyticalPotential}
The explicit forms of the pairwise potential $U(|\boldsymbol{r}|)$ used
in our simulations have been the following. 
The first case that we considered is the electrostatic interaction among 
identical molecules in  electrolytic solution; this is described by the
Debye-H\"{u}ckel potential \cite{Anderson1976}:
\begin{equation}\label{Debye}
 U_{\text{Debye}}(\boldsymbol{r})= \dfrac{(Ze)^2}{\varepsilon |\boldsymbol{r}|} \cdot
 \dfrac{e^{-\frac{2R}{\lambda_D}\left( \frac{|\boldsymbol{r}|}{2R} - 1 \right)}}{
 \left(1+ R/\lambda_D \right)^2}
\end{equation}
where $\lambda_D$ is the Debye length of the electrolytic solution, $R$ is the
molecular radius, $e$ is the elementary charge and $\varepsilon$ is the static
dielectric constant of the medium. As water is ubiquitous in
microscopic biological systems, we put
$\varepsilon=\varepsilon_{water}\simeq 80$, \textit{i.e.} its static value at room
temperature. Coulomb screening is an essential feature of biological systems
which shortens the range of electrostatic interactions due to small ions freely
moving in the environment.
In order to study how the diffusion and dynamical properties of the system
change by varying  the spatial range of the interactions, we consider different
values for $\lambda_D$ and, in the ideal case of $\lambda_{D}\rightarrow +\infty$,
 we adopt the pure Coulomb potential
for charged particles in a dielectric medium:
\begin{equation}\label{Coulomb}
 U_{\text{Coul}}(\boldsymbol{r})= \frac{(Ze)^2}{\varepsilon |\boldsymbol{r}|}
\end{equation}
The second case concerns a long-range attractive dipolar potential 
\cite{Preto2012, Preto2013,Preto2014}.
This, in regularized form, reads as
\begin{equation}
\label{dipolar_potential}
 U_{\text{Dipolar}}(\boldsymbol{r}) = - \frac{c}{|\boldsymbol{r}|^3+\alpha}
\end{equation}
where $c$ is a positive parameter and $\alpha$ is a parameter that prevents
$U(r)$ from becoming singular. This potential describes both an attractive
electrostatic and an attractive electrodynamic dipole-dipole interaction. 
In describing a system with a strong Debye shielding, the use of the potential
of Eq.\eqref{dipolar_potential} is equivalent to the
implicit assumption that this potential is of \textit{electrodynamic}
origin.
The parameter $\alpha$ flattens $U(r)$ at short distances when these are comparable with
the radius $R$ of the molecules.
In fact, when
$r$ is small, multipole moments could play a role and,
in principle, this would lead to the description of the interaction among
complex bodies whose charge distributions should be taken into
account \cite{Stone2008}.
 Here it is assumed that the
net result of these  interactions (which can be attractive as well as
repulsive), occurring  when the molecules are close one to the other, is zero.
The softened potential Eq.\eqref{dipolar_potential} solves this problem.  The parameter $\alpha$ is fixed by the condition that
the second derivative of $U$ (where the force intensity reaches its maximal
value) vanishes, that is $\alpha = 2r^3$, at $\boldsymbol{r} =0.1
\mu\text{m} $.
The value of the coefficient $c$, which controls the force intensity, has been
determined by the requirement that $U(\boldsymbol{r})$, at the same value
$\boldsymbol{r}=0.1 \mu \text{m}$, is equal to a given fraction of $-k_B T$ ,
whence $U( \boldsymbol{r}=0.1\ \mu\text{m}) =-k_B T/10$.

\subsection{Numerical algorithms}\label{numAlg}

We have numerically studied systems of $N$ molecules confined in a cubic volume of size $L$.
To get rid of spurious boundary effects, 
periodic boundary conditions (PBC) have been assumed which implies the existence
of an infinite number of replicas/images throughout the space.
As we are interested in studying dynamical properties and diffusive behavior of
different concentrations of molecules, we fixed the number of molecules $N$
and varied the average intermolecular distance $\langle d \rangle$ according to
the relation
\begin{equation}
L=\sqrt[3]{N}\langle d \rangle
\end{equation}
In presence of long-range interactions and PBC, each molecule contained in the
previously mentioned box interacts with all the molecules contained in the
above mentioned images/replicas, that is, the pairwise potential
$U(\boldsymbol r_i,\boldsymbol r_j)=U(|\boldsymbol r_i-\boldsymbol r_j)|$ 
in Eqs. \eqref{lang.} and \eqref{lang2.} has to be replaced by an
effective potential $U^{\mathrm{eff}}(\boldsymbol r_i,\boldsymbol r_j)$ of the form:
\begin{equation}
U^{\mathrm{eff}}(\boldsymbol{r}_i,\boldsymbol{r}_j)=\sum_{\boldsymbol{k}\in\mathbb{Z}^3}U(|\boldsymbol{r}_i-\boldsymbol{r}_j+\boldsymbol k L|)
\label{effpotPBC}
\end{equation}
where $\mathbb{Z}^3$ is the space of $3$-dimensional
integer vectors. In order to compute the force
$\boldsymbol F_j(\boldsymbol r_i)$ on the $i$-th particle
due to the $j$-th particles and all its
replicas, we rearrange the terms of the sum in
Eq. \eqref{effpotPBC}, so that
\begin{equation}
\begin{split}
&\boldsymbol F_j(\boldsymbol r_i)=-\boldsymbol \nabla_{\boldsymbol x_i} U(|\boldsymbol x_i-\tilde{\boldsymbol r}_j |)+\\
&+\boldsymbol \nabla_{\boldsymbol x_i}\sum_{\boldsymbol k\in \mathbb{Z}^3, \boldsymbol k\neq{\mathbf{0}}}
U(|\boldsymbol x_i-\tilde{\boldsymbol r}_j+\boldsymbol k L|)
\end{split}
\label{Forcejoni}
\end{equation}
where $\boldsymbol x_i$ is the $i$-th particle image
position into the reference box and $\tilde{\boldsymbol r}_j$
is the nearest image of $j$-th particle, that is
\begin{equation}
|\boldsymbol x_i-\tilde{\boldsymbol{r}}_j|=|\boldsymbol r_{i,j}|=\min_{\boldsymbol k\in \mathbb{Z}^3}|\boldsymbol x_i-\boldsymbol{r}_j+\boldsymbol k L|<\dfrac{L\sqrt{3}}{2}=\lambda_{NN}
\label{maxdistNN}
\end{equation}
It is clear by Eqs. \eqref{Forcejoni} and \eqref{maxdistNN} that
short and long-range interactions (in the sense specified in the Introduction)
have to be managed in two different ways.
For short range interactions it is always possible to define
a cutoff length scale $\lambda_{\text{cut}}$ such that
the effects of the interactions beyond this distance are negligible. 
In the systems we have
studied by means of numerical simulations, the Debye electrostatic
potential is a short range potential with a cutoff scale
of the order of some units of the  Debye length $\lambda_D$.
As for each case considered  it is  $\lambda_{NN}>30\lambda_{D}$,
the second term on the right-hand side of Eq.\eqref{Forcejoni} has been neglected
in numerical computations.
For long-range interactions (\textit{i.e.} Coulomb potential Eq.\eqref{Coulomb}
and dipole-dipole electrodynamic potential Eq.\eqref{dipolar_potential}),
it is not possible to define a cutoff length scale $\lambda_{\text{cut}}$
so that, in principle, the infinite sum in Eq.\eqref{Forcejoni}
should be considered. A classical way to account for long-range interactions
resorts to the so called Ewald summation \cite{Allen1989}. In the subsequent Section we describe
a more recent and practical method - replacing Ewald's one - known as Isotropic
Periodic Sum (IPS).
The equations of motion \eqref{lang2.} were numerically solved using
the Euler-Heun algorithm \cite{Burrage2007}, a
second order predictor-corrector scheme.
The position $\boldsymbol{r}_{i,n}$ of the $i$-th particle
at time $t_{n}=t_0+n\Delta t$, $t_0$ being the initial time, is obtained by:
\begin{equation}\label{heun_1}
 \boldsymbol{r}_{i,n} =  \boldsymbol{r}_{i,n-1}+\dfrac{1}{2 \gamma}\Bigr[
 \boldsymbol{F}(\boldsymbol{r}_{i,n-1})+\boldsymbol{F}
 (\tilde{\boldsymbol{r}}_{i,n}) \Bigr]\Delta t
 +\sqrt{\frac{2kT}{\gamma}}\boldsymbol\xi_{i,n-1}
\end{equation}
where $\boldsymbol{F}$ is the resultant of the forces acting on the $i$-th
particle, and  $\tilde{\boldsymbol{r}}_{i,n}$ is calculated with the Euler
predictor by:
\begin{equation}\label{heun_2}
 \tilde{\boldsymbol{r}}_{i,n}= \boldsymbol{r}_{i,n-1}+\dfrac{1}{ \gamma}
 \boldsymbol{F}(\boldsymbol{r}_{i,n-1})\Delta t
 +\sqrt{\frac{2kT}{\gamma}}\boldsymbol\xi_{i,n-1}
\end{equation}
The initial position of each particle is randomly assigned at $t_0$ using a
uniform probability distribution in a cubic box of edge $L$.\\

\subsubsection*{IPS correction to long-range potentials} 
Because of the long-range nature of Coulomb and dipolar potentials
(described by Eqs.\eqref{Coulomb} and \eqref{dipolar_potential},
respectively) the force acting on each particle is given by the sum of the
forces exerted by all the particles in the box and by the particles
belonging to the images. For the computation of these forces, we used the IPS
method \cite{Wu2005, Wu2009}, a cutoff algorithm
based on a statistical description of the images
isotropically and periodically distributed in space.
Assuming that the system is homogeneous on a length scale $R_{c}$,
we can define an effective pairwise IPS potential 
$U^{IPS}=U^{IPS}(|\boldsymbol r_{i,j}|,R_c)$ which takes into account
the sum of pair interactions within the local region and with
the images of this one:
\begin{equation}
U^{IPS}(|\boldsymbol{r}_{i,j}|,R_{c}) = \begin{cases} U(|\boldsymbol{r}_{i,j}|)+\phi(|\boldsymbol{r}_{i,j}|,R_{c}), & |\boldsymbol{r}_{i,j}|\leq R_{c} \\ \\ 0, & |\boldsymbol{r}_{i,j}|>  R_{c} \end{cases}
\end{equation}
where $\phi(|\boldsymbol{r}_{i,j}|,R_{c})$ is a correction to the potential
obtained by computing the total contribution of the interactions with
the particle images beyond  the cutoff radius $R_{c}$ \cite{Wu2005, Wu2009}.
For the Coulomb potential of Eq.\eqref{Coulomb}, we obtained an analytical
expression for the IPS correction
$\phi_{\text{Coul}}(\boldsymbol{r}_{i,j},R_{c})$.
For computational reasons this has been approximated  by a
polynomial of degree seven in $x=|\boldsymbol{r}_{i,j}|/R_c$ 
with $x$ in the interval $(0;1]$:
\begin{equation}
\label{Coulomb_IPS}
\begin{split}
\phi_{\text{Coul}}(x)= &-9.13636 \times 10^{-7}+0.000100298 x +\\
&+ 0.298588 x^2  +0.0151595 x^3 +\\
&+ 0.00881283 x^4 +  0.10849 x^5+ \\
 &-0.0930264 x^6 + 0.0482434 x^7
\end{split}
\end{equation} 
For the regularized dipole potential of Eq.\eqref{dipolar_potential}
it is not possible to compute analytically the IPS correction. Nevertheless,
since the regularization constant $\alpha$ in \eqref{dipolar_potential} could
be negligible with respect to $R_c^3$, so that $\alpha/R_c^3\ll 1$, we will
assume that the  dipolar potential has the form $U_{\text{Dipolar}}(r)\simeq c/r^3$ for
$r\geq R_c$.
Thus, we can compute the exact IPS correction
$\phi_{\text{Dipolar}}(|\boldsymbol{r}_{i,j}|,R_{c})$, and, approximating this
by means of a polynomial, we obtain:
\begin{equation}
\begin{split}
\phi_{\text{Dipolar}}^{IPS}(x)=&-3.34576\times 10^{-6 }+ 0.000199865 x +\\
&+0.936254 x^2 + 0.0259481 x^3 + \\
 &+0.0971465 x^4 + 0.184721 x^5 +\\
 &-0.146205 x^6 + 0.0877732 x^7
\end{split}
\label{DipolarIPS}
\end{equation}
We have chosen $R_c=L/2$ under the hypothesis that on this scale the system is 
 homogeneous.

\subsection{Long-time diffusion coefficient}
We aim at assessing  the experimental detectability of long-range
interactions between biomolecules taking into account 
quantities accessible by means of standard experimental techniques.
A valid approach to do so
is the study of transport properties.
For this reason, in our simulations we chose the long-time diffusion coefficient
$D$  as main observable of the system described by Eqs. \eqref{lang2.}.
This coefficient is defined, consistently with Einstein's relation
\cite{Allen1989}, as:
\begin{equation}
D=\lim_{t\rightarrow +\infty}\dfrac{\langle |\Delta\boldsymbol{r}_i(t)|^2\rangle}{6t}
\label{Dsdef2}
\end{equation}
 $\Delta \mathbf{r}_i(t)=\boldsymbol{r}_i(t)-\boldsymbol{r}_i(0)$ being
the total displacement of a particle in  space and $\langle  a_{i}\rangle=1/N\sum_{i=1}^N a_{i}$, 
the average over the particle set. We remark that in our system the displacements $\Delta \boldsymbol{r}_i (t)$
 are not mutually independent due to the interaction potential $U(|\boldsymbol{r}_i-\boldsymbol{r}_j|)$ in Eqs. \eqref{lang2.} which 
establishes a coupling between different particles; in that case, the average 
over particles index concerns correlated stochastic variables. 
Nevertheless, as our system is non-linear
with more than three degrees of freedom, it is expected to be chaotic so
that, in this case, the statistical independence of
particle motions is recovered. Moreover, when a chaotic diffusion gives  
$\langle |\Delta\boldsymbol{r}_i(t)|^2\rangle\propto t$ (which is the case of
the models considered in the present work), the diffusion coefficient $D$ is
readily   computed through a linear regression of $\langle|\Delta
\boldsymbol{r}_i (t)|^2\rangle$ expressed as a function of time.
In what follows we refer to $\langle |\Delta \boldsymbol{r}_i (t)|^2\rangle$ as
Mean Square Displacement (MSD).

%
%




\subsection{Self-diffusion coefficient for interacting
particles}\label{Sec-Virial} 
In this Section,  we  derive a formula which corrects the Brownian diffusion coefficient
 by taking into account molecular interactions described by $U(r)$ in
Eqs.\eqref{lang.}.  
Following the classical derivation given by Langevin, we rewrite 
Eqs.\eqref{lang.} in terms of the displacement of each particle with respect to
its initial position: $\Delta\boldsymbol {r}_i=\boldsymbol{r}_i(t)-\boldsymbol
r_i(0)$
\begin{equation}
\begin{split}
\label{lang.delta}
{M} \frac{\D^{2} \Delta\mathbf{r}_{i}}{\D t^{2}}  = & - \gamma \frac{\D \Delta\boldsymbol{r}_{i}}{\D t}  - \sum\limits_{j=1}^{N} \nabla_{\boldsymbol{r}_i} U(\boldsymbol{r}_{i}, \boldsymbol{r}_{j})
+\\ &+\sqrt{2 \gamma k_{B} T} \boldsymbol{\xi}_i(t) \qquad \text{for } i=1,...,N
\end{split}
\end{equation}
since $\D^{n} \mathbf{r}_i/\D t^{n}=\D^{n}\Delta\mathbf{r}_i/\D t^{n}$.
Taking the scalar product with $\Delta\boldsymbol {r}_i$ of both sides, we
obtain:
\begin{equation}
\label{langevin3}
\begin{split}
&\dfrac{1}{2}M\dfrac{\D^2|\Delta \boldsymbol {r}_i|^2}{\D t^2}-M v_i^2=-\dfrac{\gamma}{2}\dfrac{\D |\Delta \boldsymbol{r}^2_i|}{\D t}+\\&-\Delta \boldsymbol{r}_i\cdot\sum_{j\neq i}^{N}\nabla_{\boldsymbol{r}_i} U(\boldsymbol{r}_{i}, \boldsymbol{r}_{j})+\sqrt{2 \gamma_A k_B T} \Delta \boldsymbol{r}_i\cdot\boldsymbol\xi_{i}(t) \\ &\qquad \text{for } i=1,...,N
\end{split}
\end{equation}
where $v_i^2=|\D\Delta\boldsymbol{r}_i/\D t|^2=|\D \boldsymbol{r}_i/\D t|^2$.
Introducing the time derivative of the square module of the total displacement
$z_i=\D |\Delta \boldsymbol{r}_i|^2/\D t$, we obtain
\begin{equation}
\begin{split}
&\dfrac{1}{2}M\dfrac{\D z_i }{\D t}-M v_i^2 =-\dfrac{\gamma}{2} z_i -\Delta \mathbf{r}_i\cdot\sum_{i \neq j}\nabla_{\mathbf{r}_i}U (\boldsymbol{r}_{i},\boldsymbol{r}_{j})+\\
&\sqrt{2 \gamma k_\text{B} T} \Delta\boldsymbol{r}_i \cdot\boldsymbol\xi_{i}(t) \qquad \text{for } i=1,...,N
\end{split} 
\label{langevinz}
\end{equation}
According to Eq.\eqref{Dsdef2} the self-diffusion coefficient $D$ can be
equivalently expressed in terms of $z_i$ as
\begin{equation}
D=\lim_{t\rightarrow+\infty}\dfrac{1}{6t}\int_{0}^{t} \dfrac{\D \langle|\Delta\boldsymbol r_i(\tau)|^2\rangle}{\D\tau}\D\tau=\lim_{t\rightarrow+\infty}\dfrac{1}{6}\langle\langle z_i\rangle\rangle
\label{Selfdiff3}
\end{equation}
where $\langle\langle \cdot \rangle\rangle$ indicates a double mean over particles and time.
Let us now apply this double averaging to Eqs.\eqref{langevinz} and
remark that $\langle\langle\Delta\boldsymbol{r}_i
\cdot\boldsymbol\xi_{i}(t)\rangle\rangle=0$ because the time average is
equivalent to an average over noise realizations (see
Eq.\eqref{statnoise}). Thus we get:
\begin{equation}
\begin{split}
&\langle\langle z_i \rangle\rangle=-\dfrac{1}{\gamma}M\Bigr\langle\Bigr\langle\dfrac{\D z_i}{\D t}\Bigr\rangle\Bigr\rangle+\\
&+\dfrac{2}{\gamma} \Biggr[M \langle\langle v_i^2 \rangle\rangle-\langle\langle\Delta\mathbf{r}_i\cdot\sum_{i \neq j}\nabla_{\boldsymbol{r}_i}U (\boldsymbol{r}_{i},\boldsymbol{r}_{j})\rangle\rangle\Biggr]
\label{averagedzeq}
\end{split}
\end{equation}
whose limit for $t\rightarrow+\infty$  gives an expression
for the diffusion coefficient which explicitly depends on $U(r)$, according
to Eq.\eqref{Selfdiff3}.
We assume that such a limit is finite for every
term on the right hand side in Eq.\eqref{averagedzeq} and that:
\begin{equation}
\lim_{t\rightarrow+\infty}\left\langle \left\langle\dfrac{\D z_i}{\D t}\right\rangle\right\rangle=0
\end{equation}
which amounts to considering that the motion is diffusive. 
Since we consider  systems at thermodynamic equilibrium, the Equipartition
Theorem entails $\lim_{t\rightarrow+\infty}M\langle\langle v_i^2
\rangle\rangle=3k_{B}T$.
We thus obtain the following expression for the diffusion coefficient $D$
\begin{equation}
D=\lim_{t\rightarrow+\infty}D_0\left[1-\dfrac{\langle\langle\Delta\boldsymbol{r}_i(t)\cdot\sum_{i \neq j}\nabla_{\mathbf{r}_i}U (\boldsymbol{r}_{i},\boldsymbol{r}_{j})\rangle\rangle}{3 k_{B}T}\right]
\label{VirialD}
\end{equation}
where $D_0=k_{B} T/\gamma$ is the Brownian diffusion coefficient.\\
We remark that the correction term does not depend on initial conditions,
as it would appear at a first glance at the equation above.
In fact, having assumed thermal equilibrium, the dynamics is 
 self-averaging so that time averages of observables 
for very long time $t$ (ideally $t\rightarrow+\infty$) its equivalent to an
average over initial conditions \footnote{A naive computation, neglecting the effect of PBC,
would always give a value of diffusion coefficient that is increased with respect
to the Brownian one in the case of repulsive interactions, and decreased in the case of
attractive interactions. The presence of infinite replicas due to PBC makes this
statement incorrect in our case, as it can be seen using the form of the effective potential in Eq.\eqref{effpotPBC}.}.
For numerical calculations, the potential-dependent term in Eq.\eqref{VirialD} is
computed using:
\begin{equation}
\dfrac{\Delta D}{D_0}=\dfrac{D_0-D_s}{D_0}=\dfrac{1}{N}\sum_{i=1}^N\biggr(\dfrac{1}{m}\sum\limits_{k=1}^m \Delta \boldsymbol{r}_i(k\Delta t)\cdot\boldsymbol{F}_i(k\Delta t)\biggr)
\label{VirialProcedure}
\end{equation}
where $\Delta \boldsymbol{r}_i(k\Delta t)=\boldsymbol{r}_i(k\Delta t)-\boldsymbol{r}_i(0)$ is the total displacement
of the $i$-th particle at $k$-th integration step (taking into account PBC
according to Eq.\eqref{Forcejoni} and possibly IPS corrections) and
$\boldsymbol{F}_i(k\Delta t)$ is the resultant force acting on the $i$-th
particle .

\subsection{Measuring chaos in dynamical systems with noise}\label{numLyap}
Equations \eqref{lang2.} are a system of non linear differential 
equations  with additive noise. A relevant observable measuring the degree of
instability  of the dynamics is the Largest Lyapunov Exponent (LLE). 
The definition and numerical computation of the LLE is standard for noiseless
 deterministic maps and dynamical systems
\cite{Benettin1976}, while it is more debated and controversial for randomly
perturbed dynamical systems, the difficulty being due to the non differentiable
character of stochastic perturbations
\cite{Loreto1996,Grorud1996,Arnold1988}.
However, note that our system is in principle a smooth dynamical system because 
the \textit{stochastic} term in Eqs. \eqref{lang2.} 
is just a simplified way to represent the \textit{deterministic} (and
differentiable) collisional interactions between Brownian solute particles with
solvent molecules (water).
In other words  Eqs.\eqref{lang2.} are a practical representation of the
dynamical system described by the following smooth ODEs:
\begin{equation}
\gamma \frac{d \boldsymbol{r}_{i}}{dt} = - \sum\limits_{j=1}^{n}\boldsymbol{\nabla}_{\boldsymbol{r}_i} U\left(|\boldsymbol{r}_{i} -\boldsymbol{r}_{j}|\right)
+ \sqrt{2 \gamma k_{B}T}\boldsymbol f_i(t)
\label{determlyapeq}
\end{equation}
where $\boldsymbol {f}(t)=\left(\boldsymbol f_1(t),...,\boldsymbol
f_N(t)\right)$ is a $3N$-dimensional time-dependent vector of functions
representing the effect of collisions of water molecules with Brownian particles on a microscopic scale. 
If we look at  $\boldsymbol f(t)$ on a timescale comparable to the
characteristic collision time of
water molecules with Brownian particles ($\tau_{coll}\sim 1 \
\mathrm{p}\text{s}$), $\boldsymbol f(t)$ is a differentiable function 
and its Fourier spectrum has a-priori a cut-off frequency.
In spite of this, since we study the dynamics on  timescales which outnumber
$\tau_{coll}$ by at least six orders of magnitude, $\boldsymbol f(t)$ can be
safely approximated  by the standard white noise
specified  by Eqs.\eqref{whitenoise} and \eqref{statnoise}.
The white noise approach is useful for the numerical computation
of the dynamics, but the underlying physics
is in principle well described by  the ODEs system of
Eqs.\eqref{determlyapeq}. Having this in mind, we get rid of the subtleties  of
defining chaos in randomly perturbed dynamical systems and we resort to standard
computational methods \cite{Pettini2007}. Deterministic
chaos stems from two basic ingredients: stretching and folding of phase
space trajectories.
In our case the folding of trajectories in phase space is guaranteed by
PBC which make phase space compact, while stretching is given by  the local
instability of the trajectories. Hence their average instability 
 is measured through the usual
Largest Lyapunov Exponent $\lambda$, defined as:
\begin{equation}
\lambda=\lim_{t\rightarrow+\infty}\dfrac{1}{t}\ln\dfrac{\|\boldsymbol\zeta (t)\|}{\|\boldsymbol \zeta(0)\|}
\label{Lyapunovformula}
\end{equation}
where $\|\cdot\|$ is the euclidean norm in $\mathbb{R}^{3N}$ and
$\boldsymbol \zeta=(\zeta_1,...\zeta_{3N})$ is
a $3$N-dimensional vector whose time evolution is given by the following 
tangent dynamics equations:
\begin{equation}
\dfrac{\D \mathbf{\zeta}_i}{\D t}=-\dfrac{1}{\gamma}\sum_{k=1}^N\dfrac{\partial^2 U}{\partial x_i \partial x_{k}}\Biggr|_{\boldsymbol x (t)}\mathbf{\zeta}_k (t) \qquad i=1,...,3N
\label{tangentdyn}
\end{equation}
Of course, a positive LLE indicates deterministic chaos.
Using the above definition we expect that the LLE
vanishes in the absence of an interaction potential $U(r)$ 
in Eqs.\eqref{determlyapeq} since the tangent dynamics equations
\eqref{tangentdyn} becomes trivial.
Note that the term $\boldsymbol f(t)$
does not contribute to Eqs.\eqref{tangentdyn} which means
that the precise functional form of "noise" has no influence on
the chaotic properties of the system.
Besides its theoretical interest,  computing LLEs has to do also 
with the possibility, at least in principle, of working out these
quantities from experimental data. This could provide an additional observable
to probe the presence of long-range intermolecular interactions.
For numerical computations of the LLE  Eq.\eqref{Lyapunovformula} is replaced
by:
\begin{equation}
\lambda=\dfrac{1}{N_{\text{step}}\Delta t}\sum_{m=1}^{N_{\text{step}}}\ln\dfrac{\|\boldsymbol \zeta_{m}\|}{\|\boldsymbol \zeta_{m-1}\|}
\label{discretetimeLyap}
\end{equation}
where $N_{\text{step}}$ is the total number of integration steps
and $\Delta t$ is the time step.
In practice, to compute the time evolution of the tangent vector in
Eqs.\eqref{tangentdyn} for $N=1200$ particles
(consequently  for $3N=3600$ degrees of freedom) amounts to computing
about $6.5$ millions of matrix elements of the Hessian of $U(r)$ for each time.
This would be a very heavy  computational task, thus  
  we resorted to an old algorithm  described in the celebrated paper \cite{Benettin1976}.
This consists of considering a reference trajectory $\mathbf{x}(t)$
and of  computing  very short segments of varied trajectories
$\tilde{\mathbf{x}}(t)$ issuing  very close to this reference trajectory.
Details are given in the quoted paper.

\section{Numerical Results}\label{NumRes}
In the present Section we report
 the effect of long-distance interactions on
the diffusion behavior of a collection of molecules by analysing how $D$
deviates from its Brownian value.  
The numerical integration of Eqs. \eqref{lang2.} was performed using the model
potentials given in Section \ref{AnalyticalPotential}, using  the
integration algorithm with  periodic boundary
conditions, and the IPS corrections  to the interactions both described in
Section \ref{numAlg}.
The  computer code used  was written in
 Fortran$90$,  developed in a parallel computing
environment. This program was run on a computer cluster for typical durations of
$500$ to $1500$ hours (total CPU time) for each simulation. The overall CPU time
needed for the results reported in this Section amounts to about $200000$
CPU hours. All the simulations were performed considering a system of $1200$
molecules (since we typically used $120$ processors) of radius $R= 0.002 \mu\text{m}$, 
at a temperature of $300 \text{K}$, with an integration time step  $h=0.001 \mu\text{s}$ 
and each computation consisted of $5-8 \times 10^6$ steps.
In this paper, we use the following system of units:  $\mu\text{m}$ for
lengths,  $\text{kDa}$ for masses ($1\text{kDa}=6.0221 \times 
10^{-20}\text{gr}$ ) and  $\mu\text{s}$ for time.
The values of the self-diffusion coefficient $D$ have been obtained by means of
a least squares fit of the time dependence of the MSD, that is, using the
following fitting function:
\begin{equation}
\langle r^2 (t) \rangle =b_0+6D t
\label{DiffusionCoefficientProcedure}
\end{equation}
where the additive offset $b_0$ has no physical relevance,
but has been included in order to better estimate the long time
behavior of the MSD.
In the following Sections, the values of $D$ will be plotted normalized by the 
Brownian diffusion coefficient $D_0$. This coefficient is known a-priori and is
compared with the numerical outcome obtained for very low concentrations. 
These values are found to be in very good agreement within typical statistical
errors of the order of $1/\sqrt{N}=1/\sqrt{1200}$.
As we will see in the following, in addition to the standard source of diffusion
represented by the random forces $\sqrt{2 \gamma k_{B} T}
\boldsymbol{\xi}_i(t)$, another source of diffusion is given by the intrinsic
chaoticity of the particle dynamics stemming from the interparticle interactions.
The latter contribution to diffusion does not alter the linear time dependence
of the MSD. This circumstance is not new and has been reported in many examples
of chaotic diffusion \cite{Pettini1988,Ottaviani1991,Osborne1986,
Osborne1990,Crisanti1991}.
To give  a measure of spatial correlation in 
the simulated system we calculated the radial distribution function $g(r_n)$ defined as:
\begin{equation}
g(r_n)=\dfrac{1}{N}\sum_{i=1}^N\left[\dfrac{\mathcal{N}_{i,r_n}}{\dfrac{4\pi}{3}(n^3-(n-1)^3)\rho\delta^3}\right]\enskip
n=1,...,N_{Bin}
\label{radial_distribution}
\end{equation}
where $\mathcal{N}_{i,r_n}$ represents the number of particles
at an "effective" distance $r\in[r_n-\delta;r_n+\delta)$ from the
 $i$-th particle (\textit{i.e.} taking into account also different images of the system for PBC),
with $\delta=L/(2N_{Bin})$, $r_n=(2n-1)\delta$ and $\rho=N/ L^3 $. Although the
function $g(r_n)$ has a discrete domain, we will refer to it as $g(r)$ for the sake of simplicity
and as we set $N_{Bin}=1000$. We calculated the distance between all
pairs of molecules and binned them into an histogram normalized to the density
of the system. This function gives  a measure of the spatial correlation in the
system since it is proportional to the probability of
finding a molecule at a given distance $r$ from another one.
In addition we have measured the Lyapunov exponent, according to
what is given in Section \ref{numLyap}, and the correction to the
Brownian value $D_0$, according to Eq.\eqref{VirialProcedure}.

\subsection{Excluded volume effects}\label{Excluded volume effects}

As we already said, we aim at investigating the different
possible sources of deviation from Brownian diffusion, thus we begin with the
most simple possibility: excluded volume effects at the foreseen experimental
conditions. 
  \begin{figure}[h!] \centering
\includegraphics[scale=0.10,keepaspectratio=true]
{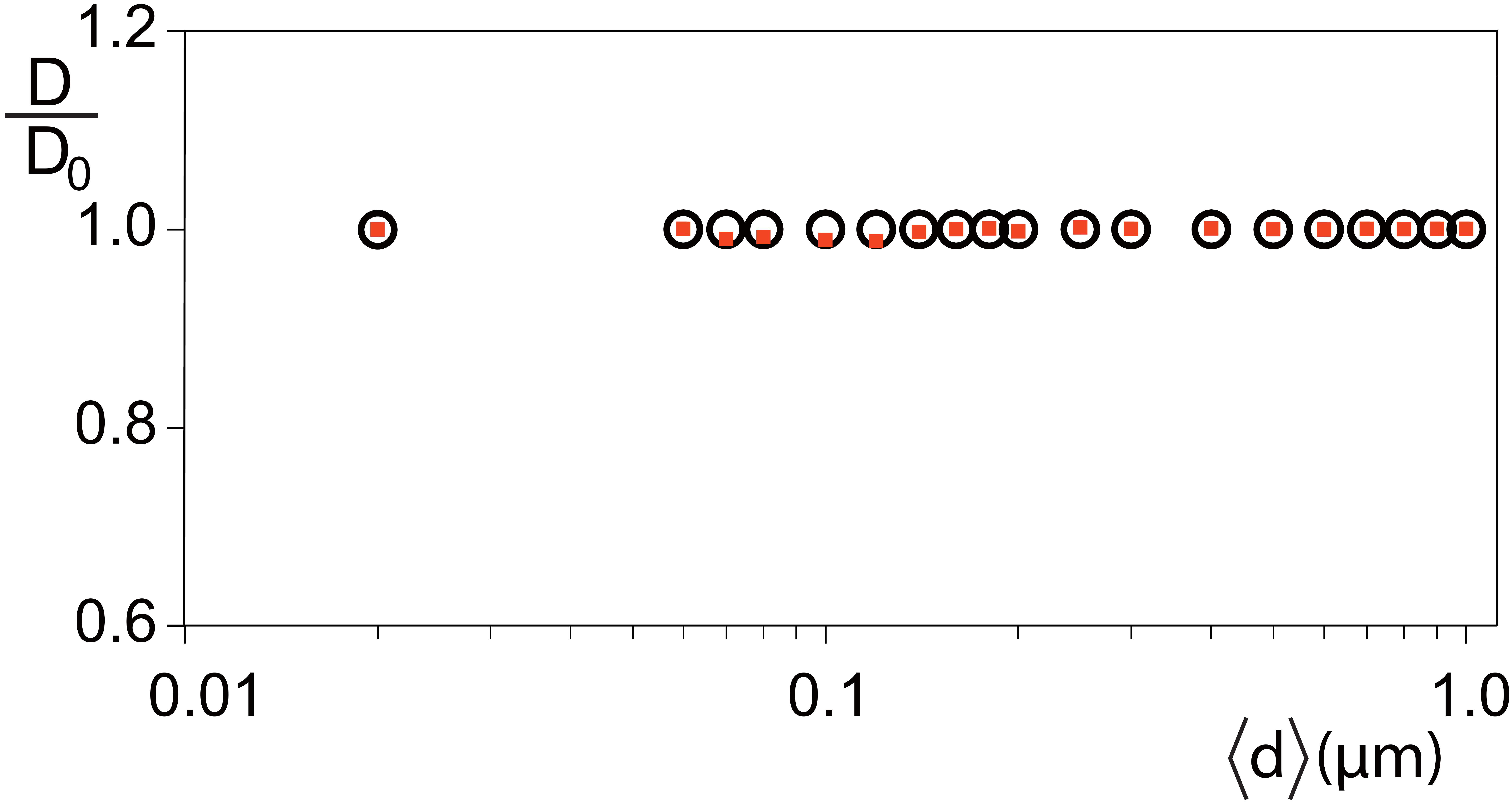} \caption[10 pt]{(Color online)  Excluded volume
simulations. Semi-log plot of the normalized theoretical self
diffusion coefficient $D/D_0$ (open circles) 
computed according to \cite{Yoshida1985} compared to the
outcomes of the standard computation (filled squares) given by
[Eqs.\protect{\eqref{Dsdef2}} and \protect{\eqref{DiffusionCoefficientProcedure}}] versus
the average distance between the particles with vanishing intermolecular
potential.}
\label{1}
\end{figure}
We considered hard-spheres with vanishing intermolecular
potential, $U = 0$, and modeling  impenetrability as follows:
whenever two molecules $i$ and $j$  get in touch and interpenetrate at some
time $t$ (that is $|\boldsymbol{r}_{i}(t)-\boldsymbol{r}_{j}(t)| < 2R$, with $R$ 
the radius of each molecule) we get back to $t-h$ and redraw the $\boldsymbol{\xi}_{i}(t)$ 
until $\boldsymbol{r}_{i,j}(t)$ are such that the impenetrability condition is
satisfied. In Figure \ref{1} we can see that the excluded volume effects on
diffusion coefficient $D$ normalized with the Brownian value $D_0$ are very
small.  These results agree with the theoretically predicted values
\cite{Yoshida1985} according to which $D=D_0 [1-2 \phi]$ where $\phi=1/6 \pi R^3 n$
and  $n=N/L^3$ is the number density.



\subsection{Effects of long and short range electrostatic interactions at fixed 
average intermolecular distance}\label{Sec-CoulombOnlyD} 
\begin{figure}[t!] \centering
\includegraphics[scale=0.11,keepaspectratio=true]{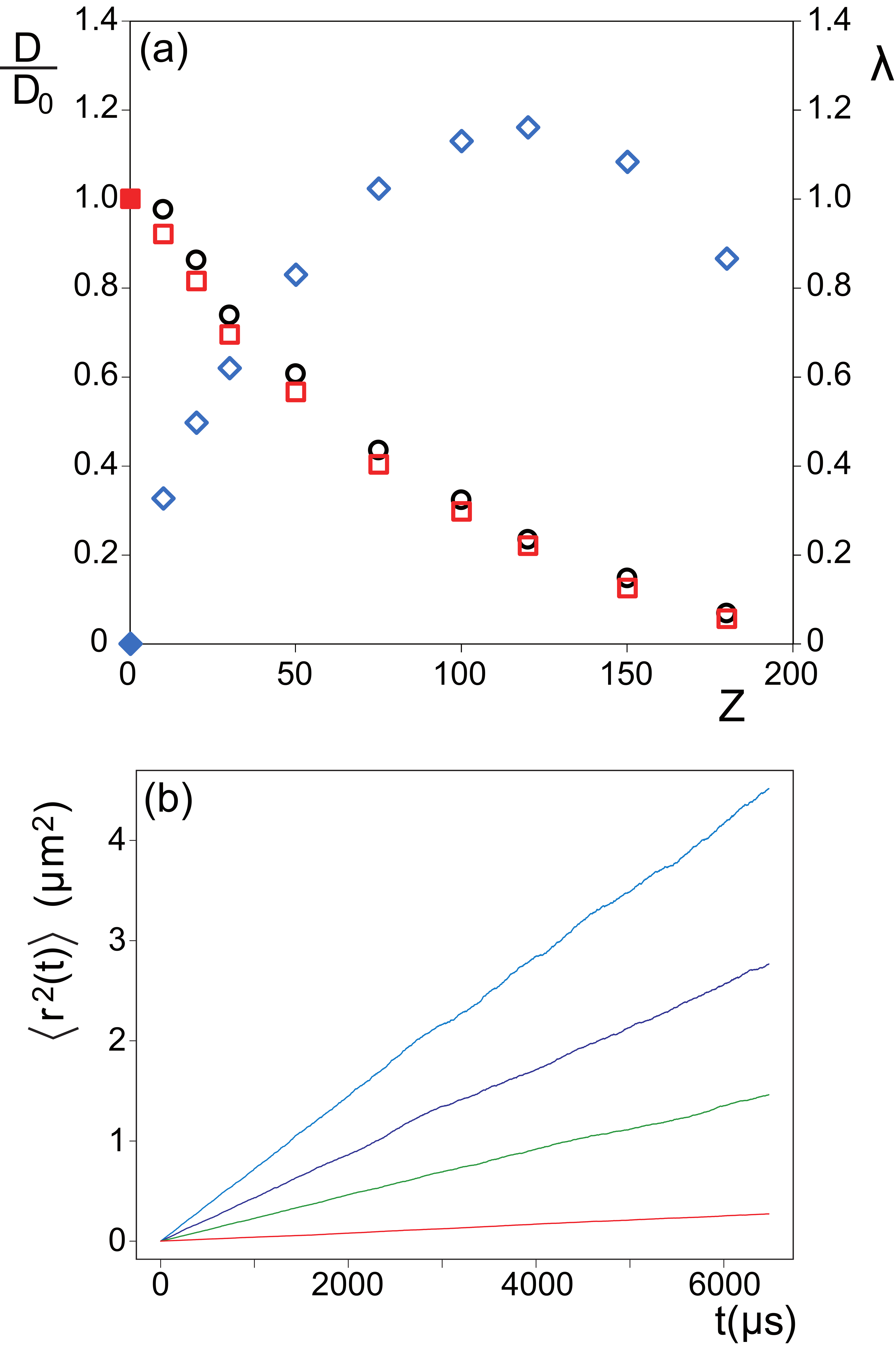}
\caption[10 pt]{(Color online) (a) Normalized self
diffusion coefficient $D/D_0$ (circles) computed according to
[Eq.\protect{\eqref{VirialProcedure}}] compared to the outcomes of the standard
computation (squares) according to [Eqs.\protect{\eqref{Dsdef2}} and
\protect{\eqref{DiffusionCoefficientProcedure}}] versus the number of charges $Z$ of the  particles interacting through Coulomb potential
with  $\lambda_D = 0.01 \mu\text{m}$ [Eq.\protect{\eqref{Debye}}] at average interparticle 
distance $\langle d\rangle=0.04\mu\text{m}$. On the second axes we report the largest
Lyapunov exponent [Eq.\protect{\eqref{discretetimeLyap}}] (rhombus).
Full symbols represent the corresponding theoretical values for vanishing $Z$
value. (b) Plot of the time evolution of the simulated MSD for different values
of charge. The charge $Z$ increases starting from the top line that corresponds
at $Z=10$, passing to $Z=50$, to $Z=100$, up to the bottom line corresponding at
$Z=180$.}
\label{Coulomb10-2-AB}
\end{figure}

\begin{figure}[t!]
\includegraphics[scale=0.115,keepaspectratio=true]{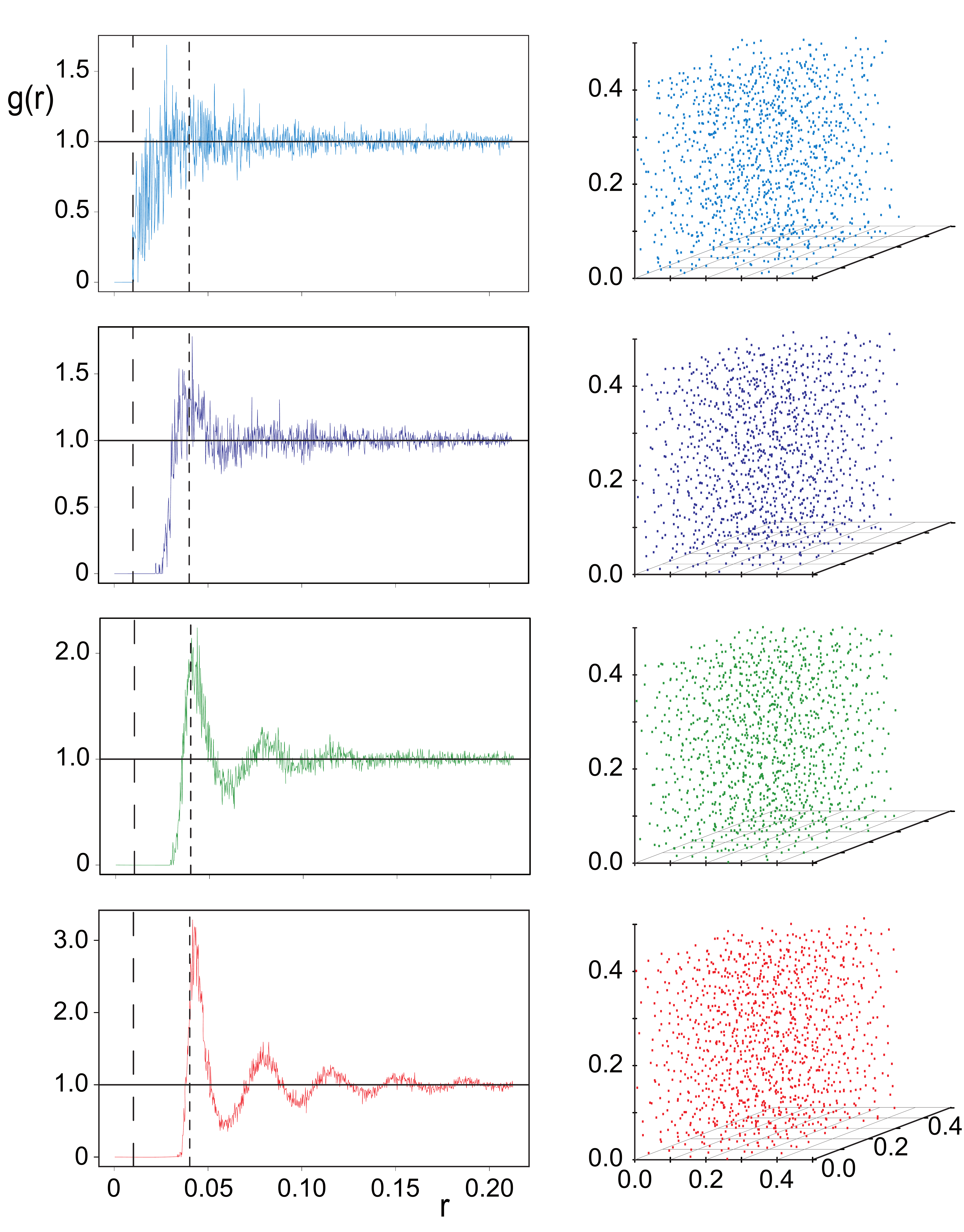}
\caption[10 pt]{(Color online) Radial distribution function $g(r)$
[Eq.\protect{\eqref{radial_distribution}}] and particles position snapshots 
at the final simulation time  for four charge values of
Figure \ref{Coulomb10-2-AB} starting from the top with $Z=10$ on the first
line, $Z=50$ on the second line, $Z=100$ on the third line and $Z=180$ on the
last one. Large dashed black line correspond to $r = \lambda_D = 0.01 \mu\text{m}$
while short dashed black line correspond to the $r = \langle d\rangle = 0.04
\mu\text{m}$. Full black line show the value $g(r)=1$. In the left panels the units of $r$ are $\mu\text{m}$, as well as
the units of the snapshots axes to the right.}
\label{Coulomb10-2-GDR}
\end{figure}

The next step is obtained by switching on interparticle interactions, keeping
fixed all the parameters (temperature, viscosity, average interparticle
distance, Debye length) but the number of charges $Z$.
This way, we can vary only the intensity of the interparticle forces measuring
the largest Lyapunov exponent and how $D$ deviates from Brownian motion. 
To begin with, the screened Coulomb potentials defined in Eq.\eqref{Debye} have been considered
for an average intermolecular distance $\langle d \rangle=0.04\mu\text{m}$ and a Debye length
$\lambda_D=0.01 \mu\text{m}$. In Figure \ref{Coulomb10-2-AB} and in Figure
\ref{Coulomb10-2-GDR}, we report the outcomes of these numerical simulations. 

 In Figure \ref{Coulomb10-2-AB}(a) we can see that the stronger the
interparticle interaction the larger the deviation from the Brownian
diffusion, that is stronger decrease of the diffusion coefficient $D$. The
 degree of chaoticity, represented by the largest Lyapunov exponent, is
also affected by the strength of the interparticle interaction.
 At the same time, the time dependence of the MSD remains linear,
that is, the chaotic diffusion still follows the Einstein-Fick law \cite{Crisanti1991},
as it can be seen in Figure \ref{Coulomb10-2-AB}(b).  
The decreasing of the diffusion coefficient occurring in presence of repulsive interactions is due
to the fact that the molecules uniformly fill all the accessible volume, thus, since there is no room for 
a free expansion of the system, the motion of any given molecule is somewhat hindered and slowed down by the surrounding ones. 
On the contrary, in presence of repulsive
forces an increase of diffusion is expected  when measured by  mutual diffusion
coefficient \cite{Tracy1992}. The latter describes the decay  of a concentration
fluctuation and it is intuitive that under the action of repulsive forces
a local higher density of particle diffuses faster than a
Brownian diffusion.
We can also observe a strikingly
good agreement between the values of $D$ obtained through the time dependence
of the MSD and by computing the theoretical corrections to Brownian value $D_0$
due to deterministic forces, according to Equation \eqref{VirialProcedure}.
The behavior of the Lyapunov exponents (Figure \ref{Coulomb10-2-AB}(a)) is characterized by an
initial increase of the chaoticity of the system with a bending - towards lower values - beginning around $Z=120$.
Such results can be qualitatively understood with the aid of the radial
distribution functions $g(r)$ reported in Figure \ref{Coulomb10-2-GDR}. 
The higher $Z$, the larger the range of spatial ordering as indicated by a
larger numbers of peaks displayed by the function $g(r)$ at distance
values which are multiples of the average intermolecular distance.
The pattern of $g(r)$ with peaks oscillating around $1$ is characteristic of
a liquid and we can observe a transition from a gaseous-like state of the
system for $Z=10$, to a short range order between $Z=50$ and $Z=100$, up to a long-range order for $Z=180$.
We can surmise that the behavior of the LLE is due to the competition between
the chaotic dynamics and the spatial ordering. To better elucidate this phenomenology, we
have considered the unscreened Coulomb potential.

\begin{figure}[t!] \centering
\includegraphics[scale=0.105,keepaspectratio=true]{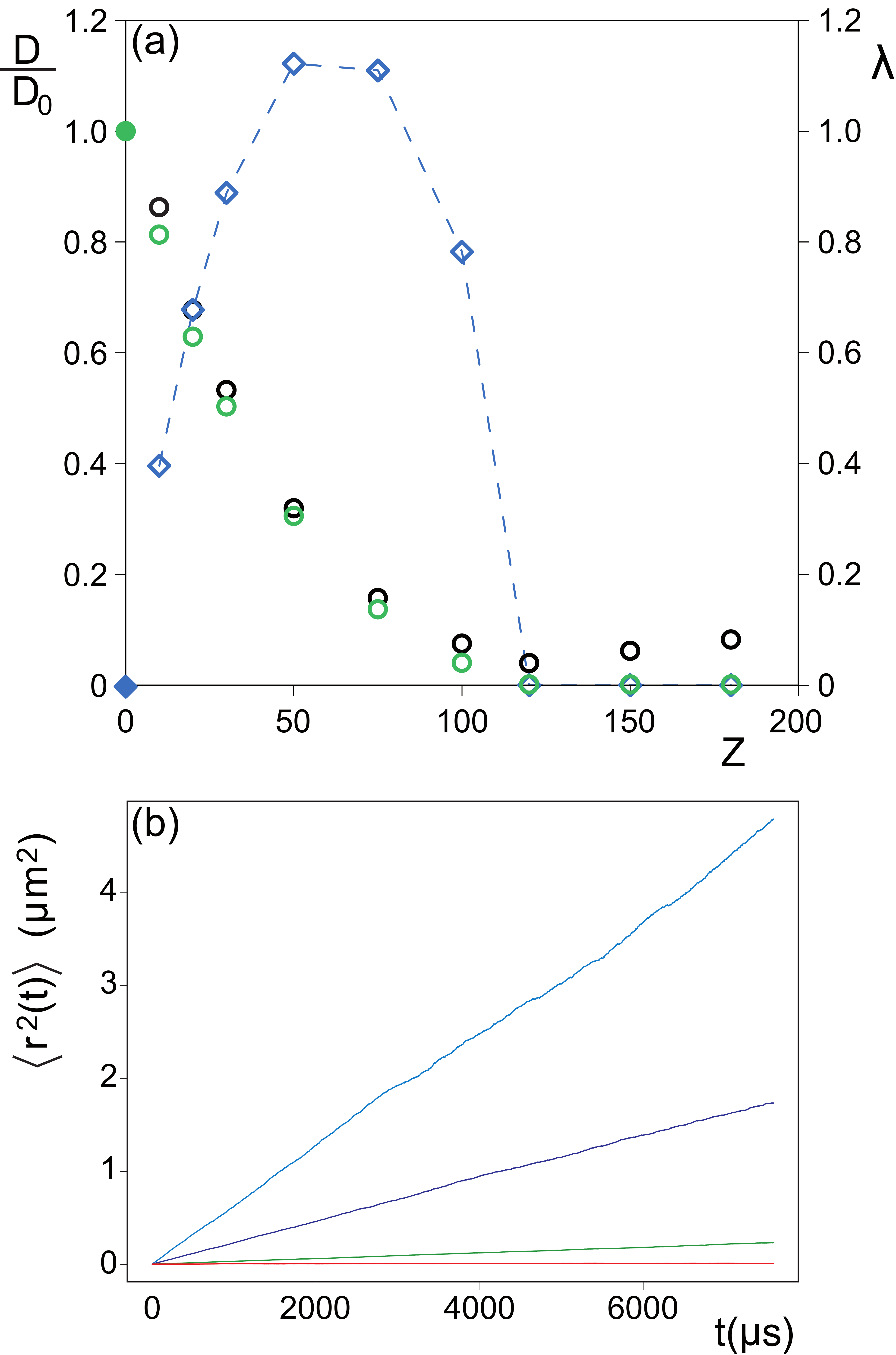}
\caption[10 pt]{(Color online) (a) Normalized self
diffusion coefficient $D/D_0$ (black circles) computed according to
[Eq.\protect{\eqref{VirialProcedure}}] compared to the outcomes of the standard
computation (grey/green circles) according to [Eqs.\protect{\eqref{Dsdef2}} and
\protect{\eqref{DiffusionCoefficientProcedure}}] versus the number of charges
$Z$ of the  particles interacting through a pure Coulomb potential
[Eq.\protect{\eqref{Coulomb}}] at average interparticle distance $\langle d\rangle=0.04\mu\text{m}$. 
On the second axes we report the largest
Lyapunov exponent [Eq.\protect{\eqref{discretetimeLyap}}] (rhombus).
Full symbols represent the corresponding theoretical values for vanishing $Z$
value. (b) Plot of the time evolution of the simulated MSD for different values
of charge. The charge $Z$ increases starting from the top line that corresponds
at $Z=10$, passing to $Z=50$, to $Z=100$, up to the bottom line corresponding at
$Z=180$.}
\label{IPS-AB}
\end{figure}

\begin{figure}[t!] \centering
\includegraphics[scale=0.115,keepaspectratio=true]{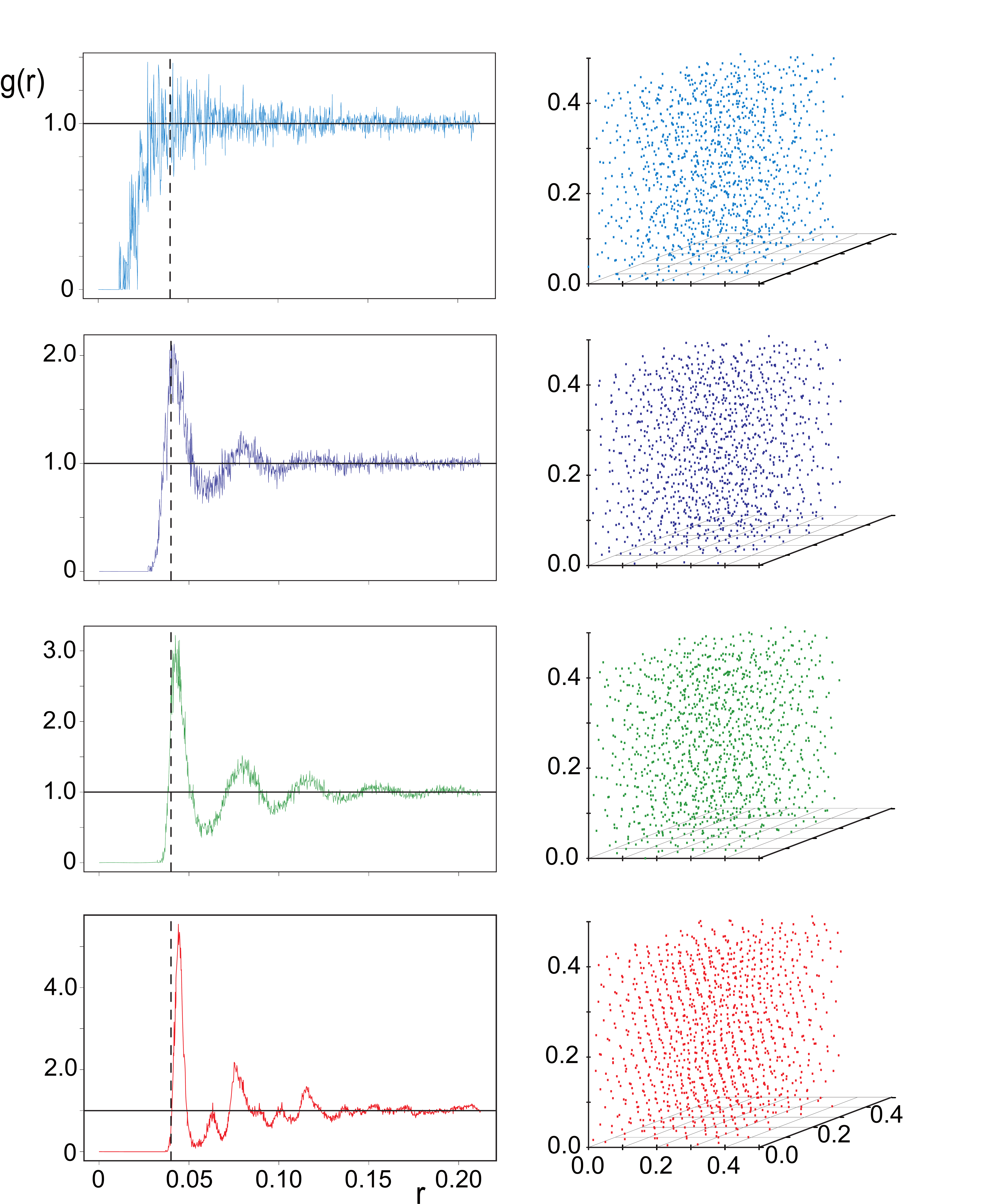}
\caption[10 pt]{(Color online)  Radial distribution function $g(r)$ [Eq.\protect{\eqref{radial_distribution}}]
and particles position snapshots at the final simulation time  for four charge
values of Figure \ref{IPS-AB} starting from the top with $Z=10$ on the first line,
 $Z=50$ on the second line, $Z=100$ on the third line and $Z=180$ on the last one.
Short dashed black line correspond to the $r = \langle d\rangle = 0.04
 \mu\text{m}$. Full black line show the value $g(r)=1$. In the left panels the units of $r$ are $\mu\text{m}$, as well as
the units of the snapshots axes to the right.}
\label{IPS-GDR}
\end{figure}

The results reported in Figures \ref{IPS-AB} and \ref{IPS-GDR} have been
obtained by means of the Coulomb potential defined in Eqs.\eqref{Coulomb} and
\eqref{Coulomb_IPS} having kept constant all the parameters (as
above with $\langle d\rangle = 0.04 \mu\text{m}$) with the
exception of the number of charges $Z$.

Likewise to Figure \ref{Coulomb10-2-AB}, we can observe that the stronger the
interparticle interaction, the larger the deviation from  Brownian diffusion, with a linear time dependence of the MSD for
all the charge values used in these simulations, as shown in Figure \ref{IPS-AB}(b).
The increase of the strength of chaos, measured by Lyapunov exponents,
observed between $Z=10$ and $Z=50$ (Figure \ref{IPS-AB}(a)) is related to the
increase of the strength of intermolecular interactions. This corresponds to a
gaseous-like state of the system as shown by the first panel of Figure \ref{IPS-GDR}.
In the second panel of the same Figure, the maximum value reached by
the LLE, at $Z=50$, is attained when a sufficient degree of spatial order sets
in so that it competes with dynamical chaos of the gaseous-like phase.
The strong decrease of the LLE observed from $Z=75$ is due to a further
enhancement of spatial order, as shown by the $g(r)$ in the third panel of Figure \ref{IPS-GDR}. 
The fourth panel of the same Figure shows a crystal-like arrangement of the
molecules confirmed by the pattern of the function $g(r)$ \cite{Allen1989}. Moreover for $Z\geq 120$ the LLE drops to
values very close to zero with a pattern displaying a seemingly sharp transition. Correspondingly, the
diffusion coefficient also drops to zero after a monotonous decrease from its Brownian value
at $Z=0$. Finally, the values of $D/D_0$ given by Eq.\eqref{VirialProcedure}, reported
in Figure \ref{IPS-AB}(a), are again in very good agreement with the outcome of
the standard computation; a growing discrepancy is observed in the above mentioned
transition occurring at $Z=120$ where the degree of chaoticity is close to vanishing.



\subsection{Effects of long and short range electrostatic interactions at fixed
charge value}\label{Sec-CoulombOnlyZ}

\begin{figure}[b!] \centering
\includegraphics[scale=0.095,keepaspectratio=true]
{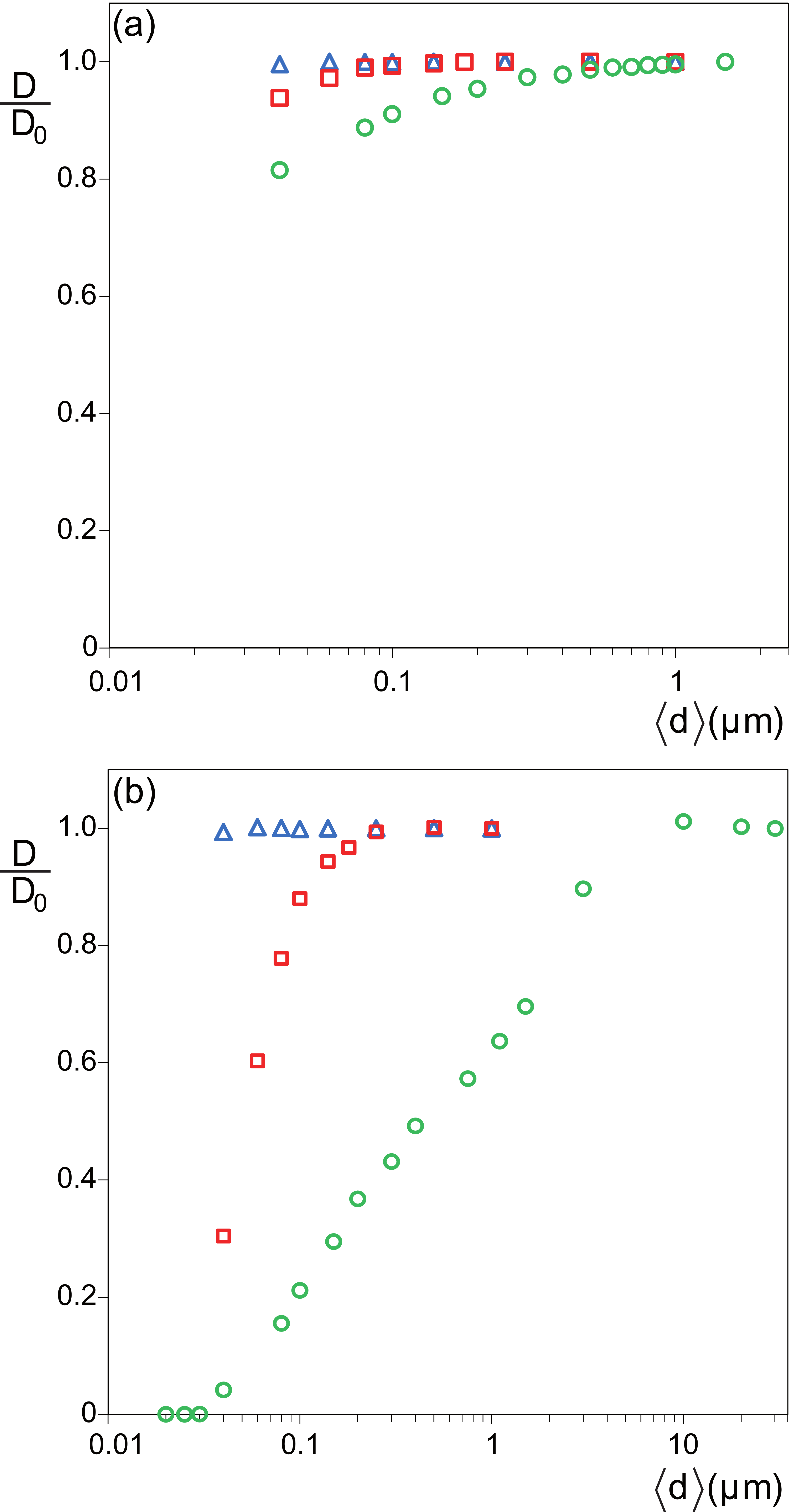} \caption[10 pt]{(Color online)  Semi-log
plot of the normalized self-diffusion coefficient $D/D_0$ versus the average distance 
of the  particles interacting through Coulomb potentials
[Eqs.\protect{\eqref{Coulomb}} and \protect{\eqref{Coulomb_IPS}}]
 for different combinations of $\lambda_D$ values at $Z=10$
(panel(a)) and $Z=100$ (panel (b)). The symbols indicate the Debye length values:
$\lambda_D = 0.001 \mu\text{m}$ correspond to triangles, $\lambda_D =
0.01 \mu\text{m}$ to  squares and $\lambda_D=\infty$ to circles.}
\label{Coul_lamdba}
\end{figure}

 \begin{figure}[b!] \centering
\includegraphics[scale=0.12,keepaspectratio=true]{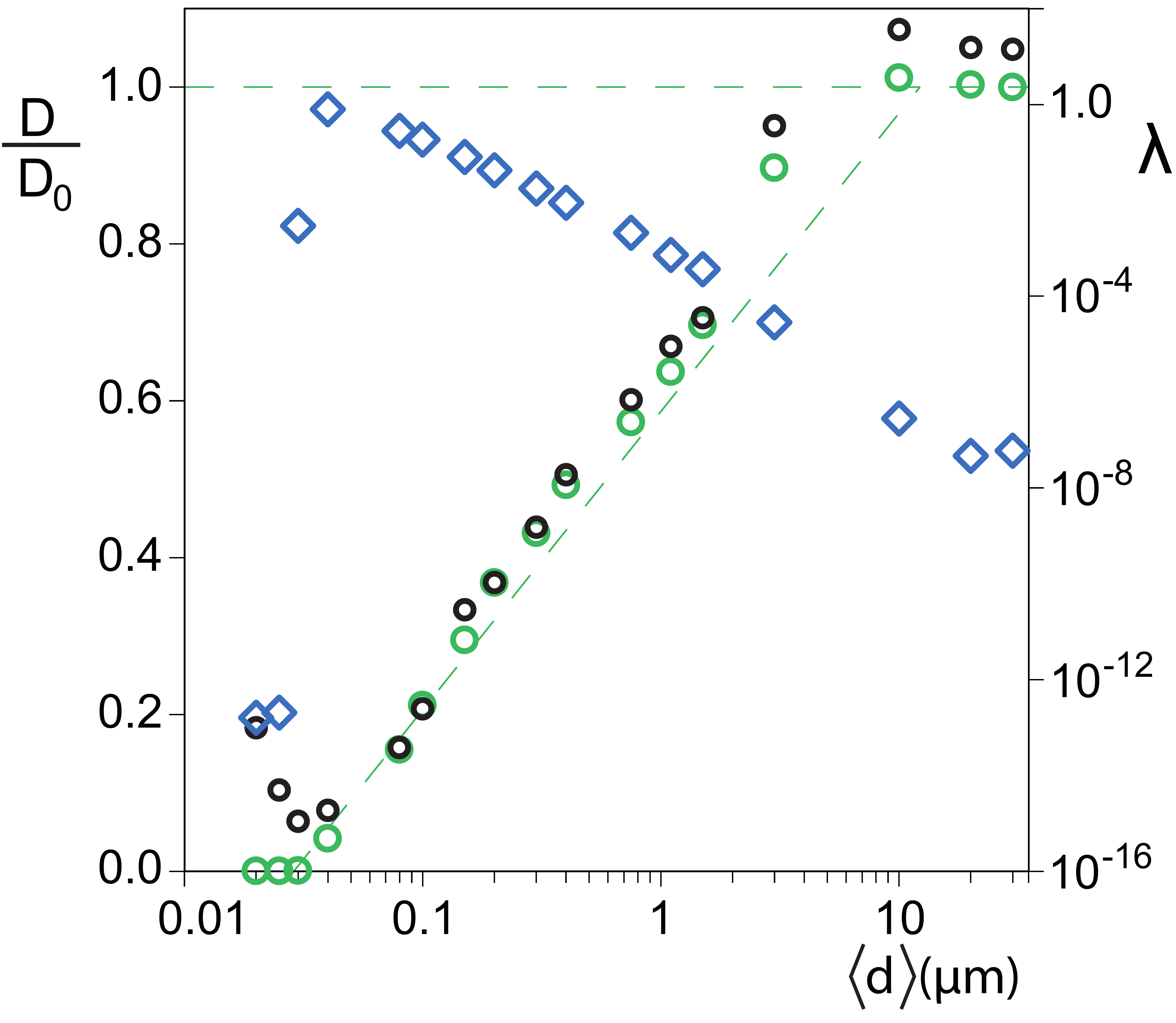}
\caption[10 pt]{(Color online)  Semi-log plot of the normalized self-diffusion coefficient versus the average distance 
of the  particles interacting through Coulomb potential
[Eqs.\protect{\eqref{Coulomb_IPS}}]  with $Z=100$. The normalized self
diffusion coefficient $D/D_0$ (black circles) computed according to
[Eq.\protect{\eqref{VirialProcedure}}] is compared to the outcomes of the
standard computation (grey/green circles) according to [Eqs.\protect{\eqref{Dsdef2}} and
\protect{\eqref{DiffusionCoefficientProcedure}}]. 
On the second axes we report the largest
Lyapunov exponent [Eq.\protect{\eqref{discretetimeLyap}}] (rhombus).
The dashed lines are guides to the eye.}
\label{IPS_Z}
\end{figure}

Let us now consider the effect of changing the interaction strength
resulting from a variation of the
average intermolecular distance and a variation of the action radius
of electrostatic forces.
This is obtained by using different Debye lengths
($\lambda_D=0.001$ and $0.01\ \mu\mathrm{m}$) for the screened Coulomb potential defined
in Eq.\eqref{Debye} and by using the Coulomb potential defined in
Eqs. \eqref{Coulomb} and \eqref{Coulomb_IPS} ($\lambda_D=\infty$), for different charge values
($Z=10$ and $Z=100$).

The choice of these parameter values is partially inspired, on the one side,
by the typical range of values of charges for proteins and for small
fragments of nucleic acids, and, on the other side, the lowest value
$\lambda_D=0.001\ \mu \mathrm{m}$ is approximately the Debye length
of the cytosol while longer Debye lengths are relevant for prospective
in vitro experiments.
Figure \ref{Coul_lamdba} summarizes the
dependence of the normalized mean diffusion coefficient  as a function of
the average distance among the molecules.
Different values of $\lambda_D$ are considered for $Z=10$ (Fig.\ref{Coul_lamdba}
(a)) and $Z=100$ (Fig.\ref{Coul_lamdba} (b)).
We can observe that at low concentrations
diffusion reaches its Brownian limit characterized by $D/D_0 \simeq 1$, and the larger the Debye length
and the number of charges, the larger the decrease of the diffusion coefficient.
It turns out that an appreciable change in the diffusion coefficient
shows up for $\lambda_D \geq 0.01 \mu\text{m}$.
The outcomes of numerical computations obtained for $Z=100$ and $\lambda_D=\infty$ are reported
also in Figure \ref{IPS_Z} and compared with the values of
the LLE and of the outcomes of the theoretical correction to the Brownian
diffusion coefficient (\eqref{VirialProcedure}).
At very high dilutions corresponding to an average interparticle distance larger
than $10\ \mu \mathrm{m}$, the diffusion is Brownian while at shorter
interparticle distances the effect of electrostatic interactions is again a decrease of
the diffusion coefficient up to a concentration corresponding to
$\langle d\rangle=0.03 \ \mu \mathrm{m}$ where diffusion stops.
By resorting to the computation of the radial distribution functions  we
observe the same phenomenology reported in Figure \ref{IPS-GDR}, that is,
in the case of Brownian diffusion the corresponding radial distribution function
closely resembles to that in first panel of Figure \ref{IPS-GDR}.
When diffusion deviates from being purely Brownian the radial distribution 
shows regular peaks as in the second and third panel of Figure \ref{IPS-GDR}
and it looks like that in the forth panel of Figure \ref{IPS-GDR}
when diffusion stops. At the same time, we observe an increase
of the LLE which corresponds to the decrease of $D$ up to the
point where $D$ vanishes. When $D$ vanishes, a sudden drop of the LLE
is observed to practically zero values. Finally, we observe a very
good agreement of the theoretical correction to the Brownian
diffusion coefficient except when diffusion stops; this suggests
that a developed chaoticity of the dynamics is a requisite for such
a computation to be reliable.

\subsection{Long range attractive dipolar effects}\label{Long range attractive dipolar effects}
\begin{figure}[t!] \centering
\includegraphics[scale=0.12,keepaspectratio=true]{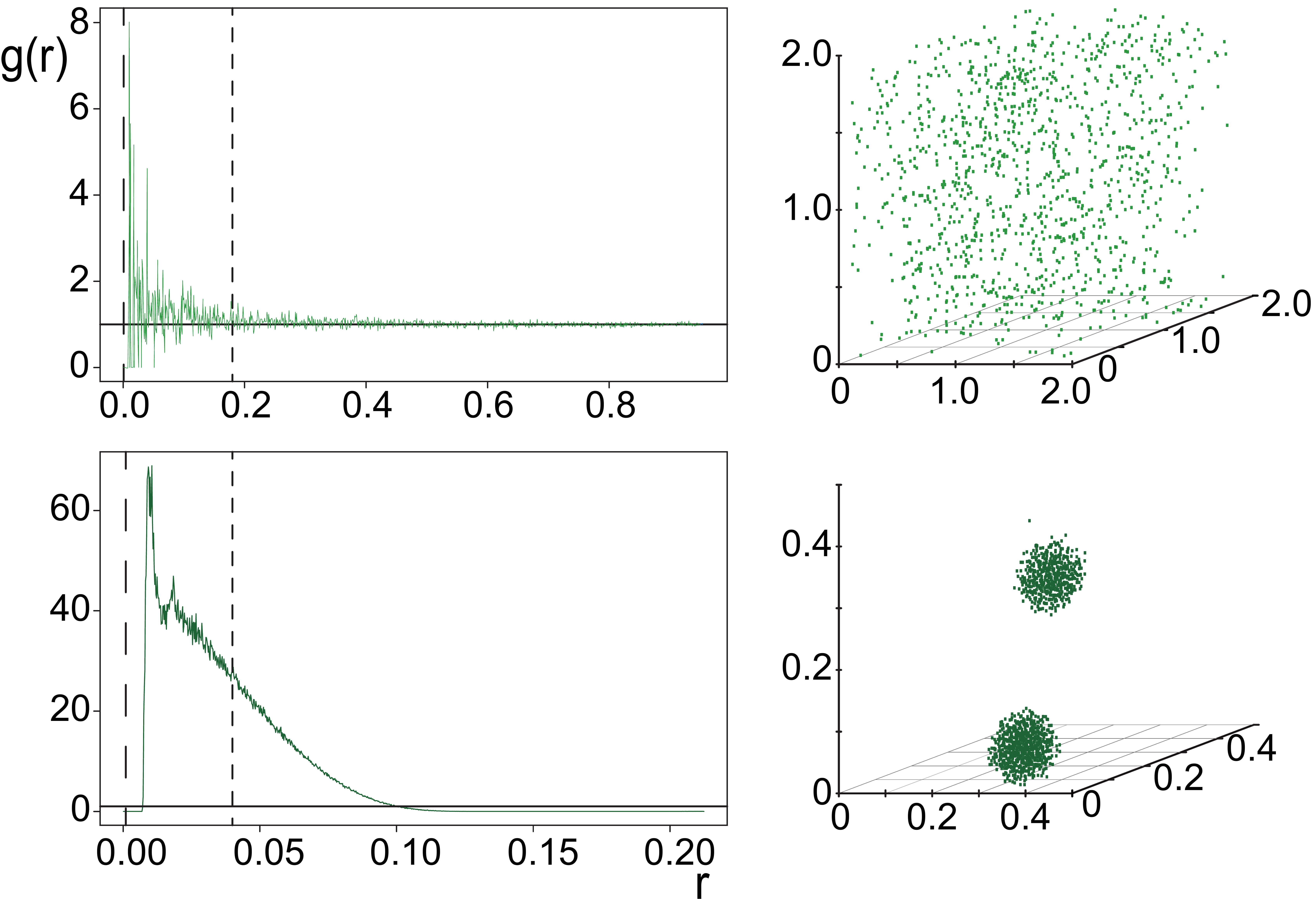}
\caption[10 pt]{(Color online) 
 Radial distribution function $g(r)$  [Eq.\protect{\eqref{radial_distribution}}] and particles
position snapshots at the final simulation time at two average interparticle
distance $\langle d\rangle=0.18\mu\text{m}$ (first line) and $\langle
d\rangle=0.04\mu\text{m}$ (second line) for particles of $Z=100$ interacting
with a Coulomb potential [Eq.\protect{\eqref{Coulomb}} with  $\lambda_D =
0.001 \mu\text{m}$] and with a dipolar potential [Eq.\protect{\eqref{DipolarIPS}}].
The large dashed black line corresponds to $r = \lambda_D$ while short dashed
black line corresponds to the $r = \langle d\rangle$. The full black line shows the
value $g(r)=1$. In the left panels the units of $r$ are $\mu\text{m}$, as well as
the units of the snapshots axes to the right.}
\label{CoulDip_GDR}
\end{figure}

As remarked in the Introduction, we are interested in verifying the experimental
detectability of long-range interactions among molecules of biological interest through their
diffusive behavior. In this Section, we focus on the study of
diffusive and dynamical properties of the system when
both electrostatic Debye potential, described in Eq.\eqref{Debye}, and
attractive dipole-dipole electrodynamic potential, described in Eqs.
\eqref{dipolar_potential} and \eqref{DipolarIPS}, are involved.
The choice of considering the simultaneous presence of these two kinds of
interactions is motivated by the fact that biomolecules are charged objects
with non-vanishing dipolar moments.
\begin{figure}[t!] \centering
\includegraphics[scale=0.095,keepaspectratio=true]
{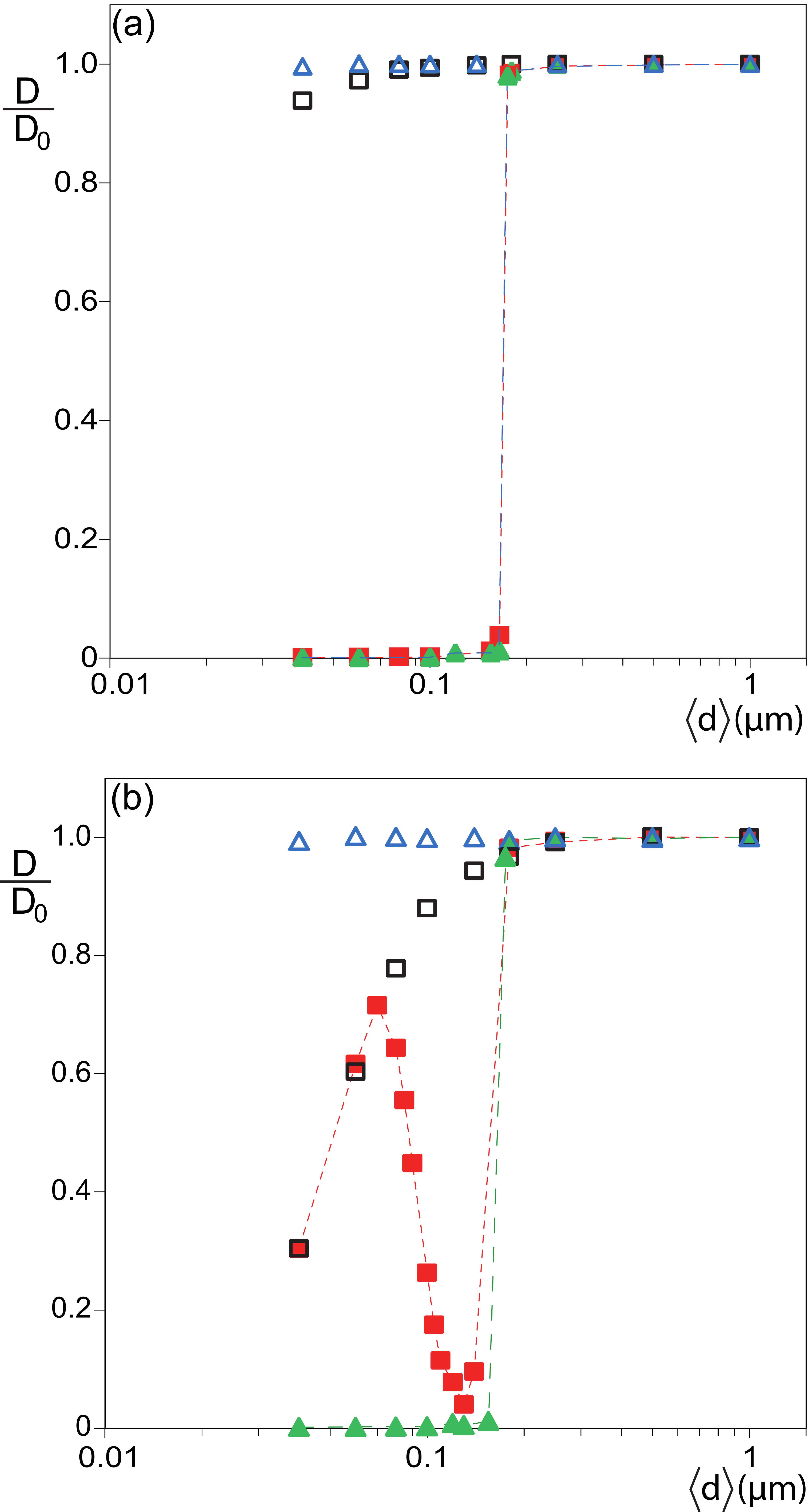} \caption[10 pt]{(Color online)  
Semi-log plot of the normalized self-diffusion coefficient $D/D_0$ versus the average distance 
of the  particles interacting only through Coulomb potential
[Eq.\protect{\eqref{Coulomb}}] and through Coulomb potential and the
attractive dipole-dipole potential [Eq.\protect{\eqref{DipolarIPS}}] for
different combinations of $\lambda_D$ values at $Z=10$ (panel (a)) and $Z=100$
(panel (b)). The symbol shapes indicate the Debye length
values, $\lambda_D = 0.001 \mu\text{m}$ correspond to triangles and
$\lambda_D = 0.01 \mu\text{m}$ to squares,  while open symbols represent
Coulomb potential and full ones the combined action of Coulomb and
dipole-dipole potentials.}
\label{CoulDip_Z}
\end{figure}
The dynamical properties and diffusive behavior in presence of an attractive
interaction are qualitatively different from those observed in the previous 
sections regarding only the repulsive Coulomb potential.
For the sake of clarity, we present and compare the combined presence
of Coulomb and dipole-dipole electrodynamic potentials (represented by full
symbols) with the  presence of only Coulomb potential (represented by open
symbols), the latter already presented in the previous Section. The kind
of symbol corresponds, as before, to the different Debye length
values: triangles correspond to $\lambda_D=0.001 \ \mu \mathrm{m}$ and squares
to $\lambda_D=0.01 \ \mu\mathrm{m}$.
In Figure \ref{CoulDip_Z} the numerical outcomes for the  normalized
diffusion coefficient, $D/D_0$, are reported as a
function of the average intermolecular distance  for two charge values, $Z=10$
(Fig. \ref{CoulDip_Z}(a)) and $Z=100$ (Fig. \ref{CoulDip_Z}(b)) and different
values of the Debye lengths, both in presence
and in absence of dipole-dipole electrodynamic potential.
 At very high dilutions, in a range between  $\langle d \rangle = 1 \mu m$ and  $\langle d \rangle = 0.2 \mu m$ the diffusion 
follows its Brownian limit characterized by $D/D_0 \simeq 1$ for each
combination of charge or potential as observed in both panels of the 
aforementioned figure.
 Let us resume first the results when only Coulomb
potential is involved; in order to observe a significant deviation from the
Brownian limit the Debye length must be at least equal to $0.01 \ \mu
\mathrm{m}$ (open squares) with a more pronounced effect for $Z=100$ where the
deviation from Brownian motion reaches  $D/D_0 \simeq 0.3$.
To begin with, we switch on the dipolar potential focusing on the lower 
charge value,  $Z=10$ (Fig. \ref{CoulDip_Z}(a)).
 We can observe a sharp decrease of the normalized
diffusion coefficient, with a transition between a diffusive Brownian motion and
an absence of diffusion. 
\begin{figure}[t!] \centering
\includegraphics[scale=0.12,keepaspectratio=true]{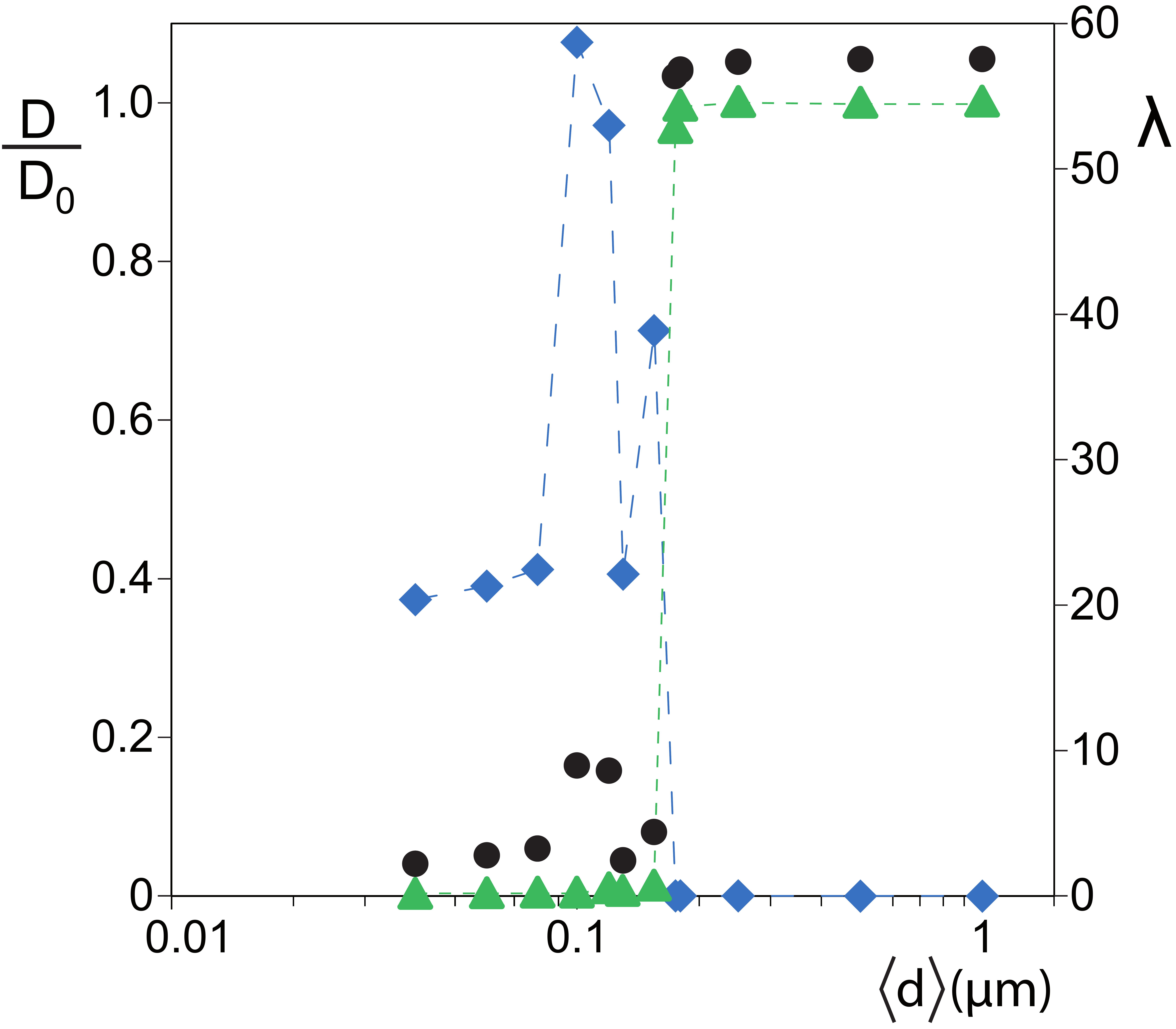}
\caption[10 pt]{(Color online) 
 Semi-log plot of the normalized self-diffusion coefficient versus the average distance 
of the  particles interacting through Coulomb and dipolar potential
[Eqs.\protect{\eqref{Coulomb}} and \protect{\eqref{DipolarIPS}}] 
with $Z=100$ and $\lambda_D = 0.001 \mu\text{m}$. The normalized self-diffusion coefficient $D/D_0$ 
(circles) computed according to
[Eq.\protect{\eqref{VirialProcedure}}] is compared to the outcomes of the
standard computation (triangles) according to [Eqs.\protect{\eqref{Dsdef2}} and
\protect{\eqref{DiffusionCoefficientProcedure}}]. On the second axes we report the largest 
Lyapunov exponent [Eq.\protect{\eqref{discretetimeLyap}}] (rhombus).}
\label{CoulDip_lyap}
\end{figure}
These results are independent of the action radius of
Coulomb potential, in fact no difference has been observed between the two
different Debye length values.
The results reported in Figure \ref{CoulDip_Z}(b)) are obtained by switching on
the dipolar potential and by increasing the intensity of Coulomb potential
(taking $Z=100$).
When the Coulomb interactions is weak ($\lambda_D=0.001 \ \mu \mathrm{m}$  full
triangle), so that the dipolar contribution overcomes it,
 we can observe the same aforementioned sharp transition
characterized by no diffusion.
On the contrary, with a larger Debye length ($\lambda_D=0.01 \
\mu\mathrm{m}$ full square)  the effects of a competition between the two
potentials, repulsive and attractive respectively, are observed when
the average intermolecular distance is varied. At large average intermolecular
distances the particle motions are  practically independent one from the other
resulting in a  Brownian diffusion, while at shorter  distances the mutual
interactions play an important role.
The interplay between the repulsive and
attractive interactions leads to a diffusion behavior
dominated by the dipolar interactions in a small range of distances in
correspondence of the transition from $D/D_0 \simeq 1$ to $D/D_0 \simeq 0$, as
it is observed in Figure \ref{CoulDip_Z}(a).
At smaller values of $\langle d\rangle$, the dipolar effect on diffusion
is balanced by the presence of short-range Coulomb repulsion, 
thus preventing the formation of a clustered system.
In Figure \ref{CoulDip_lyap}, we report the outcomes of numerical computations
of $D/D_0$ \textit{versus} $\langle d\rangle$ obtained in the case of a dominant
dipolar potential with respect to the Coulomb one ($Z=100$ and Debye
length $\lambda_D= 0.001 \ \mu m$).
In the same figure, we add to $D/D_0$, the values of the LLE and of the outcomes
of the theoretical correction to the Brownian diffusion coefficient due to
interparticle interactions (Eq. \eqref{VirialProcedure}).
This figure shows a good agreement between the theoretical correction to $D_0$
and the numerical results. We can also observe that the transition from a diffusive to
a non-diffusive behavior goes with a sharp increase of the LLE, indicating a
transition from a non-chaotic to a chaotic dynamics.
Note that, in the transition region, fluctuating patterns of the LLE and of the
theoretical correction to $D_0$ are found. We can surmise that in this region,
since the dynamics displays long transients to the final clustered
configurations, some memory of the initial conditions could be kept. 
In Figure \ref{CoulDip_GDR} the radial distribution functions of the
particles and the snapshots of their positions are given. These results refer
to two average interparticle distances and confirm a transition from
a gaseous-like state to a clustered configuration.

Finally, let us note that the results presented in the current Section
indicate a possibility to disentangle the effects of electrostatic and
electrodynamic interactions. In fact, by using a sufficiently high ion
concentration in prospective experiments, and so weakening the electrostatic
forces, only the effects of electrodynamic interactions would be
observed. 

\section{Concluding remarks}\label{Sec-concluding}

As already stated in the Introduction, the present work is the sequel of a
recent one aimed at assessing the experimental possibility of detecting 
long-range electrodynamic interactions between biomolecules. At variance with
the outcomes of the previous work, the substantial advance provided by the
present one consists of a conceptual proof of feasibility of an experimental
approach resorting to an actually measurable observable.
In particular, this observable is the diffusion coefficient that can
be measured by means of several available techniques like pulsed-field
gradient nuclear magnetic resonance
forced Rayleigh scattering (FRS), Fluorescence Recovery After Photobleaching
(FRAP) and Fluorescence Correlation Spectroscopy (FCS) to mention some of them. The long-range electrodynamic forces
we are after have been hitherto elusive to observation in spite of many studies on the diffusion
behavior of biomolecules in solution. We surmise that no evidence has been 
until now reported about the presence of these interactions because they are not
compatible with thermal equilibrium \cite{Preto2013,Preto2014}
contrary to previous predictions \cite{Frohlich1977}.
The consequence being the need for an out-of-equilibrium driving of the
biomolecules by means of a source of collective excitation.
In order to achieve the above mentioned assessment about experimental
detectability of electrodynamic intermolecular interactions, we
have performed numerical simulations whose outcomes can be summarized as
follows: \\
\textit{i)} We have found that, for dilute systems ($\langle
d\rangle$ ranging from  about $400 \mathrm{\AA}$  up to $30000 \mathrm{\AA}$ ), the
diffusion coefficient is sensitive to all the interactions considered. Starting with a uniform distribution of molecules in all the accessible volume,
an interesting phenomenon is observed: the diffusion
coefficient decreases independently of the repulsive or attractive nature of the
molecular interactions (repulsive Coulomb
with and without screening, attractive electrodynamic dipole-dipole).\\
\textit{ii)} Moreover, we observed that, in the gaseous-like phase, a decrease
of the diffusion coefficient is always accompanied by an increase of chaos.
On the contrary, when spatial order sets in, a decrease of the diffusion
coefficient is always accompanied by a decrease of chaos.
Even though it is well known that no simple relation exists between Lyapunov 
exponents and transport properties in dynamical systems, the qualitative 
correspondences observed are consistent with the intuitive idea that both 
phenomena are related to the intensity of intermolecular interactions.\\
\textit{iii)} Nice transitional phenomena have been observed: for
Coulomb interactions a first transition from purely stochastic diffusion to 
chaotic plus stochastic diffusion is found; then, at sufficiently high
concentrations, a spatial ordering of the molecules is found resembling to a
crystal-like structure. For dipole-dipole interactions an abrupt clustering
transition is observed, which is strongly reminiscent of an equilibrium
phase transition.\\
\textit{iv)} The simple theoretical model proposed in Section \ref{Sec-Virial}
gives the good values of the diffusion coefficients
computed along the dynamics in presence of intermolecular interactions
within a few percent of error.
This result paves the way - at least in principle - to analytic predictions if
the time averages used in this work are replaced by statistical averages
Eq.\eqref{gibbs_measure} worked out with the Boltzmann-Gibbs weight
Eq.\eqref{gibbsdistr} (which is the stationary measure associated with our model
equations). \\
From the experimental point of view, which was the main motivation of the present
work, we conclude that the variations of the diffusion coefficient
$D$ with respect to its Brownian value, as well as the patterns of $D$ versus
the average interparticle distance $\langle d \rangle$, are such that the
practical possibility exists of experimentally tackling the problem of interest
by means of, for example, one of the above mentioned techniques.

\begin{acknowledgments}
The authors would like to thank  J. Tuszynski and A. Vulpiani for useful comments
and discussions. The authors acknowledges the financial support of the Future 
and Emerging Technologies (FET) Program within the Seventh Framework Program 
(FP$7$) for Research of the European Commission, under the FET-Proactive grant 
agreement TOPDRIM, number FP$7$-ICT-$318121$. Pierre Ferrier laboratory is
supported by institutional grants from Inserm and CNRS, and by grants from the Commission of 
the European Communities, the 'Agence Nationale de la Recherche' (ANR), 
the 'Institut National du Cancer' (INCa), the 'ITMO Cancer from the Alliance Nationale pour les Sciences de
la Vie et de la Sant\'e (AVIESAN)'
 and the 'Fondation Princesse Grace de la Principaut\'e de Monaco'. We 
 warmly acknowledge the financial support of
 the PACA Region.
\end{acknowledgments}

\bibliographystyle{apsrev4-1}
\bibliography{ElectroMag}

\end{document}